\documentclass[useAMS,usenatbib,twocolumn]{mn2e}
\usepackage{times,amsmath,amssymb,amsfonts,latexsym}
\usepackage{multirow,graphicx,fleqn}
\usepackage{color,ulem}
       \def\msun{{{\rm M}_{\odot}}}
\def\mpch{\,{h^{-1} {\rm Mpc}}}          \def\hmpc{\,{h {\rm Mpc}^{-1}}}
\newcommand{\dd}{{\rm d}}                
             \newcommand{\dm}{{\rm dm}}
   \newcommand{\ba}{{\rm b}}
\newcommand{\vv}{{\bf v}}                \newcommand{\vk}{{\bf k}}

\newcommand{\apj}{ApJ}                   \newcommand{\apjs}{ApJS}
\newcommand{\mnras}{MNRAS}               \newcommand{\aap}{A\&A}
\newcommand{\araa}{ARA\&A}               \newcommand{\apjl}{ApJL}
\newcommand{\aj}{AJ}                     
\newcommand{\nat}{Nature}                \newcommand{\prd}{Phys. Rev. D}
\newcommand{\prl}{Phys. Rev. Lett.}      
                 \newcommand{\physrep}{Phys. Rep.}
\newcommand{\jcap}{J. of Cosmology Astropart. Phys.}
\begin{document}
\title[Deviation of baryons from dark matter]{Turbulence-induced deviation between baryonic field and dark matter field in the spatial distribution of the Universe}
\author[Yang, He, Zhu \& Feng]{Hua-Yu Yang$^1$, Ping He$^{1,2}$\thanks{E-mail: hep@jlu.edu.cn},  Weishan Zhu$^3$ and Long-Long Feng$^{3,4}$ \\
$^{1}$College of Physics, Jilin University, Changchun 130012, P.R. China. \\
$^{2}$Center for High Energy Physics, Peking University, Beijing 100871, P.R. China. \\
$^{3}$School of Physics and Astronomy, Sun Yat-Sen University, Guangzhou 510275, P.R. China. \\
$^{4}$Purple Mountain Observatory, CAS, Nanjing 210008, P.R. China.}
\date{\today}
\maketitle
% ------------------------------------------------------------------------------------------------
\begin{abstract}
The cosmic baryonic fluid at low redshifts is similar to a fully developed turbulence. In this work, we use simulation samples produced by the hybrid cosmological hydrodynamical/$N$-body code, to investigate on what scale the deviation of spatial distributions between baryons and dark matter is caused by turbulence. For this purpose, we do not include the physical processes such as star formation, supernovae (SNe) and active galactic nucleus (AGN) feedback into our code, so that the effect of turbulence heating for IGM can be exhibited to the most extent. By computing cross-correlation functions $r_{\rm m}(k)$ for the density field and $r_{\rm v}(k)$ for the velocity field of both baryons and dark matter, we find that deviations between the two matter components for both density field and velocity field, as expected, are scale-dependent. That is, the deviations are the most significant at small scales and gradually diminish on larger and larger scales. Also, the deviations are time-dependent, i.e. they become larger and larger with increasing cosmic time. The most emphasized result is that the spatial deviations between baryons and dark matter revealed by velocity field are more significant than that by density field. At $z=0$, at the $1\%$ level of deviation, the deviation scale is about $3.7\mpch$ for density field, while as large as $23\mpch$ for velocity field, a scale that falls within the weakly non-linear regime for the structure formation paradigm. Our results indicate that the effect of turbulence heating is indeed comparable to that of these processes such as SN and AGN feedback.
\end{abstract}

% ----------------------------------------------------------------------------------------------
\section{Introduction}

\label{sec:intro}

Modern cosmology reveals that the Universe is composed of three major components, i.e., dark energy, dark matter, and baryonic matter \citep{Peebles2003}. Dark energy is the dynamical mechanism that is responsible for the acceleration of the expansion of the Universe. Dark matter is believed to be composed of collisionless particles, which may be cold non-relativistic species and may collapse to form virialized dark matter halos under the interaction of gravitation \citep{Bertone2005}. Baryonic matter occupies a small fraction with only $\sim 4\%$ of the total energy content and usually exists in the form of stars and diffuse media such as intergalactic medium (IGM), or intra-cluster medium (ICM). Baryons respond to gravitational interaction and should follow the collapsing of dark matter to form high-density structures, but different from dark matter, they are collisional and hence there exist pressure effects among them, which may well be subject to thermodynamical laws. Baryons also involve much complex physical processes, such as atomic emission or absorption, IGM/ICM heating or cooling, star formation, feedback processes, chemical evolution, or turbulent motion \citep{Shu1991, Shu1992, Mo2010}. These processes may be involved into many issues, such as galaxy formation and evolution, or may induce many puzzles, such as the overcooling problem of hierarchical galaxy formation \citep{Voit2005} or the missing baryon problem \citep{Bregman2007}.

We do not know the spatial distribution of dark matter in the Universe since we are not able to detect dark matter directly via electromagnetic observations. Instead, we usually determine the distribution of dark matter indirectly through that of baryonic matter. Gravitational lensing is an effective means to reconstruct the background distribution of dark matter \citep{Schneider2006}. A series of current and upcoming weak lensing surveys, such as DES \citep{Troxel2018}, LSST \citep{Ivezic2019}, Euclid \citep{Laureijs2011} and WFIRST \citep{Green2011}, aim to measure the matter power spectrum with extremely high accuracy in order to derive accurate and precise cosmological parameter values. To fully match the accuracy of these observations, given a set of cosmological parameters, models should have to predict the non-linear matter power spectrum at the level of per cent or better for scales corresponding to comoving wavenumbers $0.1 \lesssim k \lesssim 10 \hmpc$ \citep{vanDaalen2011}.

In recent several decades, cosmological numerical simulations, both pure dark matter and hybrid hydrodynamical simulations, have become important approaches in cosmological studies \citep[e.g.][and references therein]{Frenk1999, Feng2004, Kravtsov2005, Heitmann2008}. Recent projects with hydrodynamical simulations of galaxy formation and on the relative spatial distribution of baryons and dark matter, besides gravitational clustering, have taken into account various complex processes such as cooling and heating, star formation, supernovae (SNe) and active galactic nucleus (AGN) feedback. These simulations, including OWLS \citep{Schaye2010}, cosmo-OWLS \citep{LeBrun2014}, BAHAMAS \citep{McCarthy2017,vanDaalen2020}, Illustris \citep{Vogelsberger2014}, IllustrisTNG \citep{Springel2018}, EAGLE \citep{Hellwing2016}, and Horizon \citep{Dubois2016}, have reached sufficient volume to make precision predictions for clustering on cosmologically relevant scales.

Apart from the hydrodynamical simulation projects, there are also multiple studies with analytic or semi-analytic methods, such as an optimized variant of the halo model, designed to produce accurate matter power spectra well into the non-linear regime for a wide range of cosmological models \citep{Mead2015}, or a baryonic correction model that modifies the density field of dark-matter-only $N$-body simulations to mimic the effects of baryons from any underlying adopted feedback recipe \citep{Schneider2015, Schneider2019}, or an effective field theory approach \citep{Lewandowski2015}. Compared with simulations, these approaches are much swifter to adjust relevant parameters in order to see how and to what extent the corresponding physical effects would influence the results.

A noteworthy result from these investigations is the finding that the spatial distribution of baryonic matter does not exactly follow that of dark matter. This result is strongly supported by the discovery of the so-called Bullet Cluster, from which one can find that dark matter and baryonic matter are separated in spatial distribution \citep{Markevitch2005,Clowe2006}. The finding of galaxies that are devoid of dark matter \citep{vanDokkum2018, vanDokkum2019, Guo2019} also supports the deviation between baryons and dark matter in the spatial distribution. Incidentally, these deviations between baryons and dark matter in spatial distribution are good evidence against the feasibility of alternative gravitational theories such as the modified Newtonian dynamics \citep[MOND;][]{Milgrom1983, Bekenstein2004}.

What kind of mechanisms causes these deviations? There exist some popular heating mechanisms once proposed to overcome the `cooling crisis' of hierarchical galaxy formation, i.e. overcooling problem, such as feedback processes like galactic winds from star formation and SNe, or AGN activity \citep[][and references therein]{pc11}. Simulations indicate that these heating mechanisms, especially AGN feedback, play pivotal roles to separate baryons from dark matter in spatial distribution \citep{Chisari2018}. For example, OWLS \citep{vanDaalen2011}, Illustris \citep{Vogelsberger2014}, IllustrisTNG \citep{Springel2018}, Horizon \citep{Chisari2018}, and EAGLE \citep{Hellwing2016} show that AGN feedback can have a significant impact to the total matter power spectrum at $k>10\hmpc$. Especially, OWLS and Illustris indicate that the feedback can lead to the deviation of per cent level to the total power spectrum at $k\sim1\hmpc$. For more details, we refer the interested readers to a recent review by \citet{Chisari2019} of the effects of baryons on matter clustering with cosmological hydrodynamical simulation.

In this work, however, we would like to explore another mechanism of separating baryons and dark matter rather than these baryonic feedback processes. Based on deep X-ray data and a new data analysis method for Perseus and Virgo cluster of galaxies, \citet{Zhuravleva2014} found that turbulence heating is sufficient to offset radiative cooling and indeed appears to balance it locally at each core radius of the two clusters, which indicates that turbulence heating of IGM may be a significant heating mechanism. Moreover, previous theoretical studies also revealed that the highly evolved cosmic baryonic fluid at low redshifts is similar to a fully developed turbulence \citep{Hep2006, Zhu2010, Fang2011, Zhu2013, Zhu2015, Zhu2017}, which is a support to the idea of turbulence heating for the ICM.

Turbulence heating for IGM has never been investigated by the simulations aforementioned. In this paper, we would like to explore the effects of turbulence and investigate to what extent the spatial distribution of cosmic baryons deviates from that of dark matter by turbulence heating of IGM, through the density field and velocity field by using cosmological simulations. Our simulations are performed with a hybrid dark matter/hydrodynamical code, which uses particle-mesh (PM) scheme to compute the evolution and distribution of dark matter particles, and a five-order accuracy Weighted Essentially Non-Oscillatory (WENO) scheme to compute the baryonic quantities such as temperature, density, and velocity \citep{Feng2004, Zhu2013}. Due to its five-order accuracy, this code can effectively capture shockwave and turbulence structures in the baryonic gas \citep{Zhu2015, Zhu2017}. We do not include the processes such as star formation, SN and AGN feedback into our simulations. By excluding the influence of these feedback processes, we expect that the effects of turbulence heating of IGM can be exhibited to the most extent.

This paper is organized as follows. In Section~\ref{sec:linear}, we present a linear bias model for the deviation between baryons and dark matter. In Section~\ref{sec:data}, we briefly introduce our simulation code and the data used in this work. In Section~\ref{sec:result}, we give results about the scale-dependent and environment-dependent deviation, and discussions about the mechanism of the deviation for density and velocity. In Section~\ref{sec:concl}, we present the summary and conclusions. In setting up the simulations, we adopt the Lambda cold dark matter ($\Lambda$CDM) concordance cosmological model, with the {\it WMAP5} parameters as $\Omega_{\rm m}=0.274$, $\Omega_{\Lambda}=0.726$, $h=0.705$, $\sigma_8=0.812$, $\Omega_\ba=0.0456$, and $n_s=0.96$ \citep{Komatsu2009}.

% --------------------------------------------------------------------------------------
\begin{figure*}
\centerline{
\includegraphics[width=0.45\textwidth]{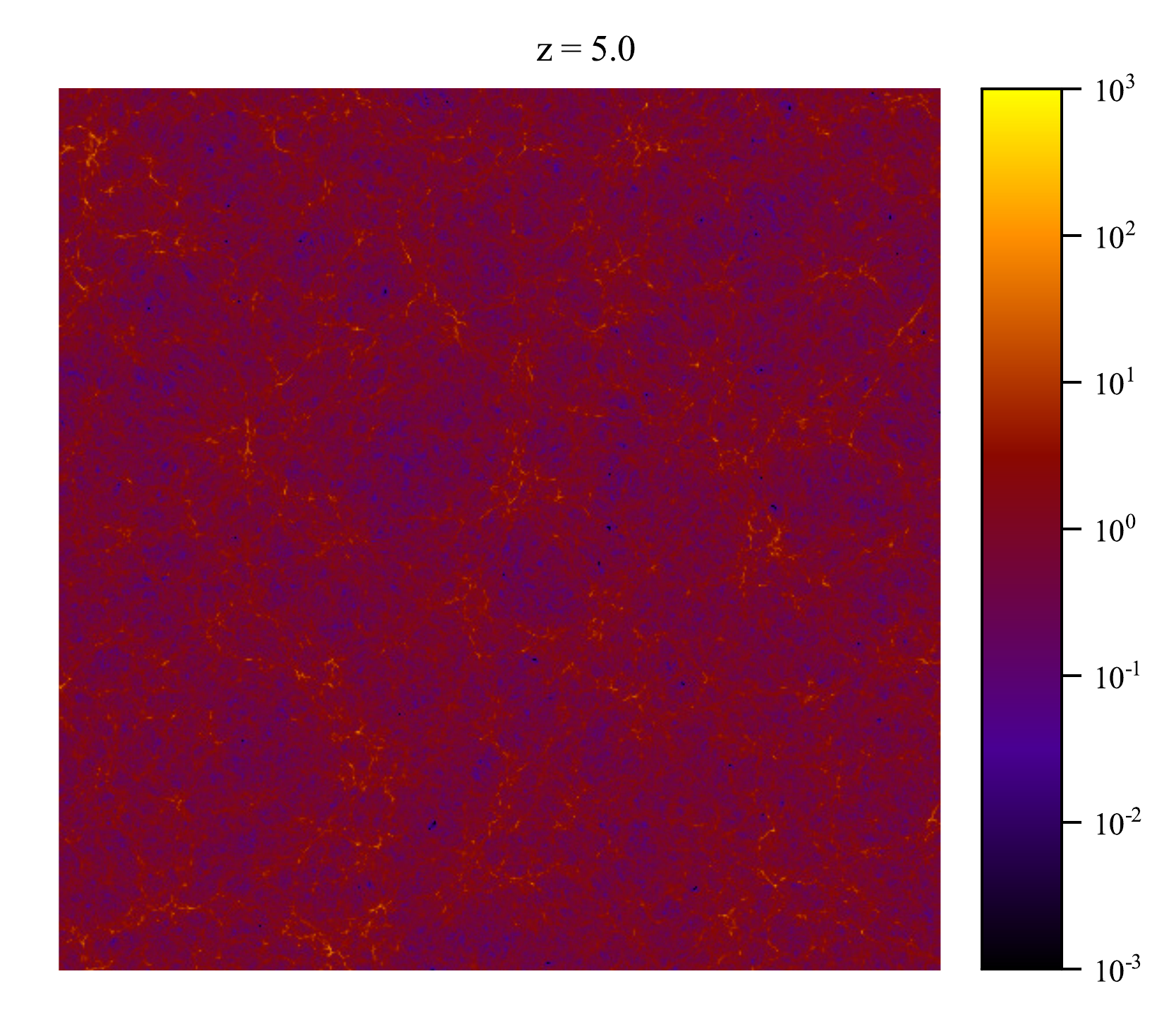}
\includegraphics[width=0.45\textwidth]{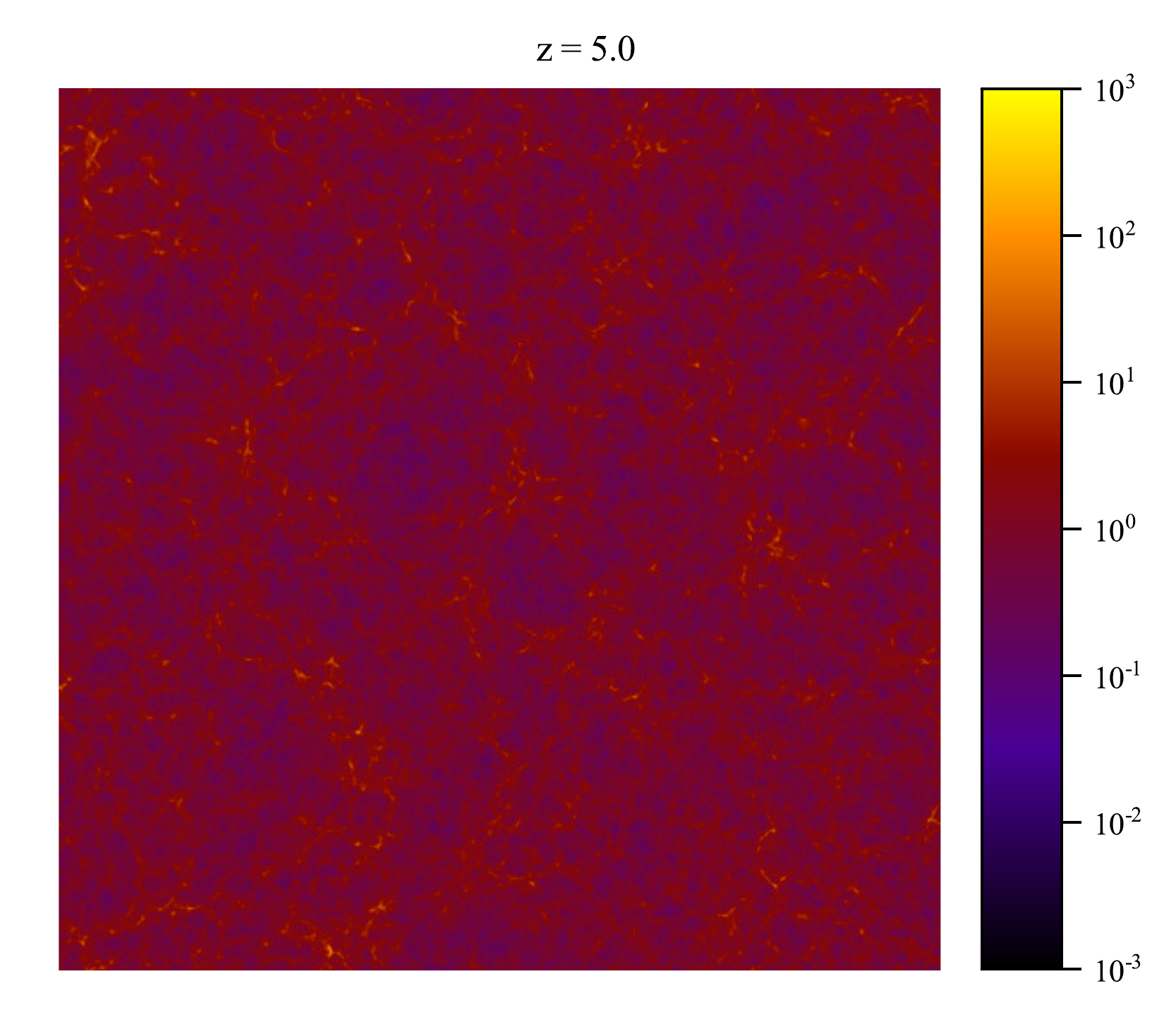}
}
\centerline{
\includegraphics[width=0.45\textwidth]{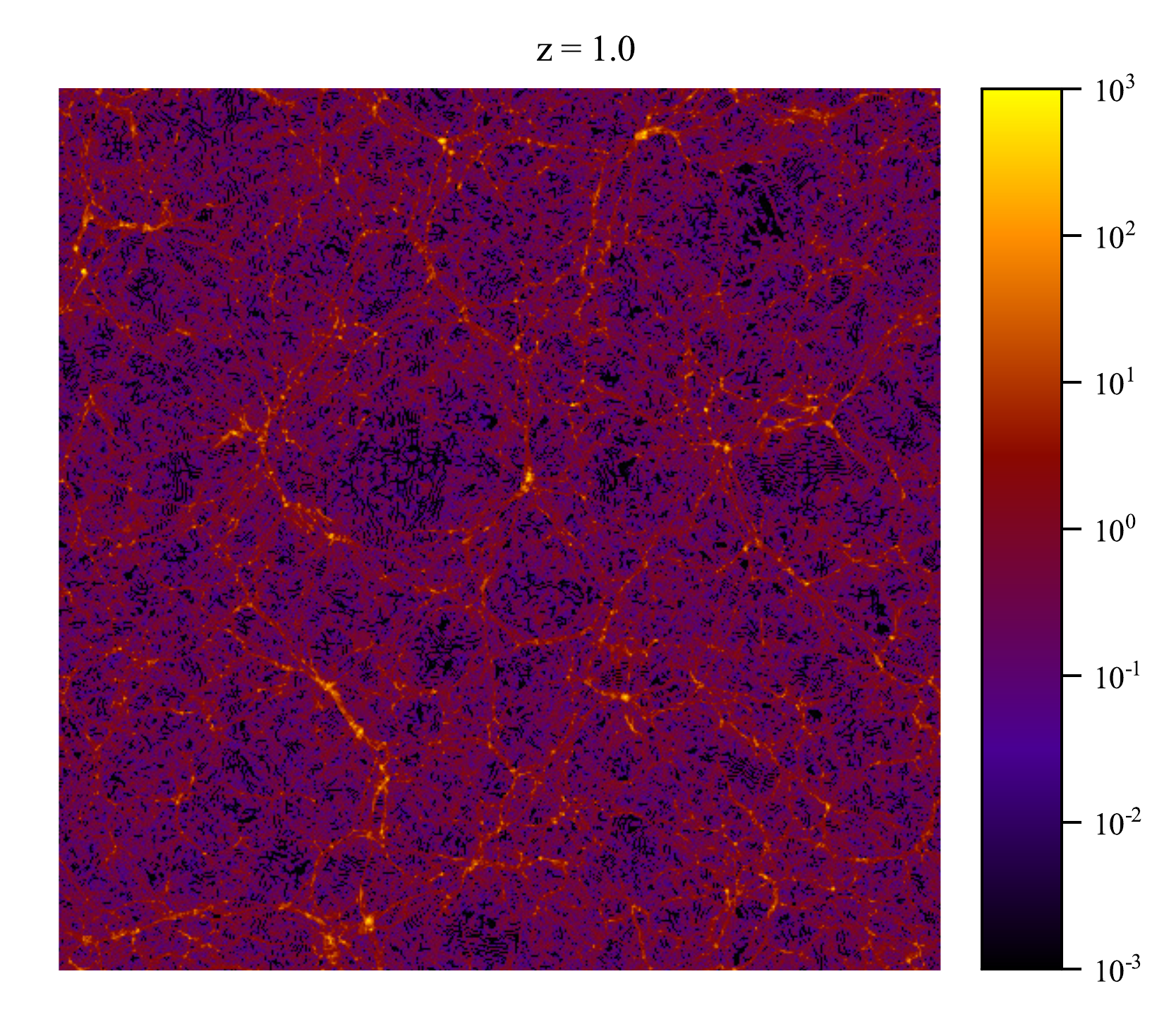}
\includegraphics[width=0.45\textwidth]{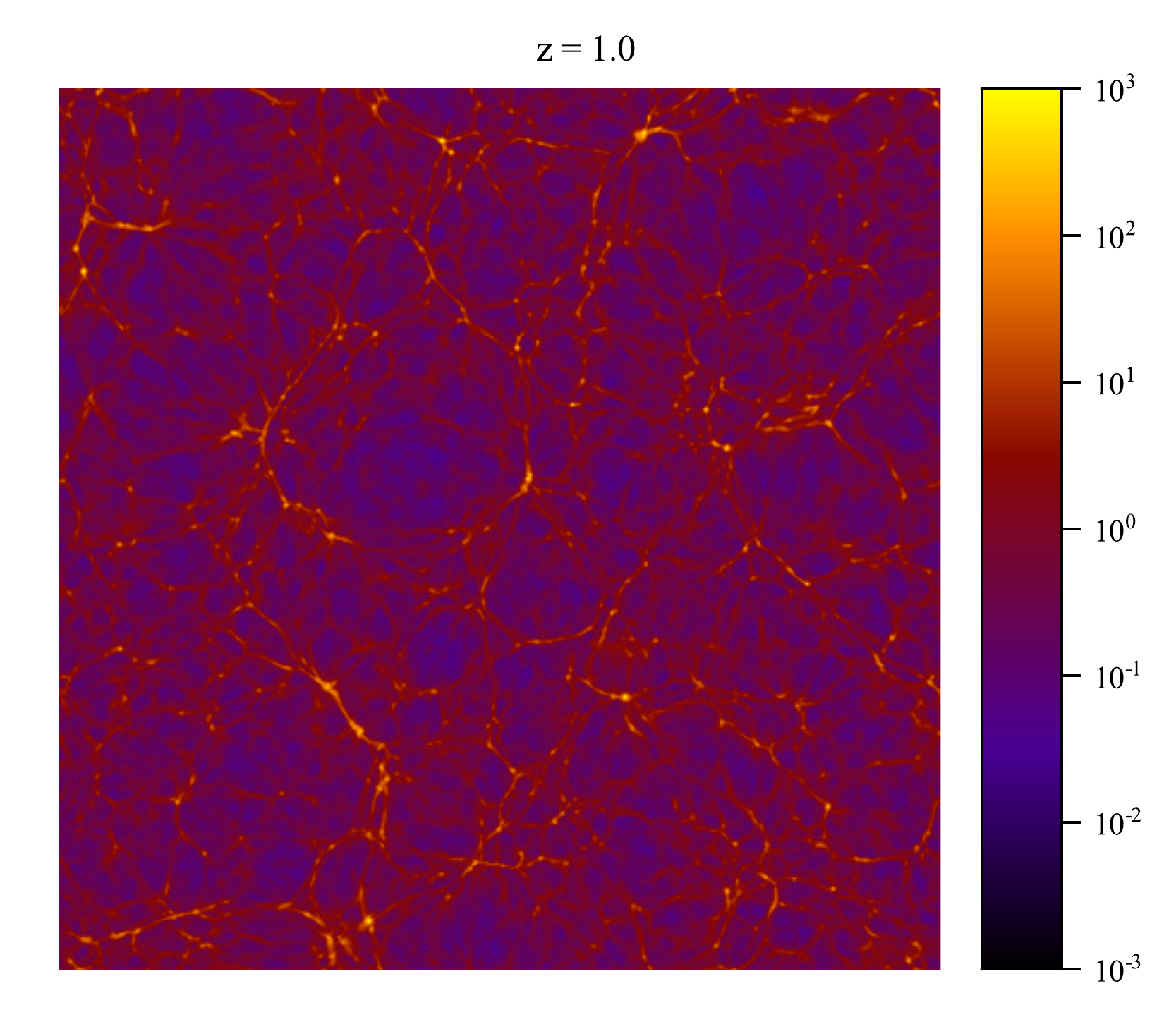}
}
\centerline{
\includegraphics[width=0.45\textwidth]{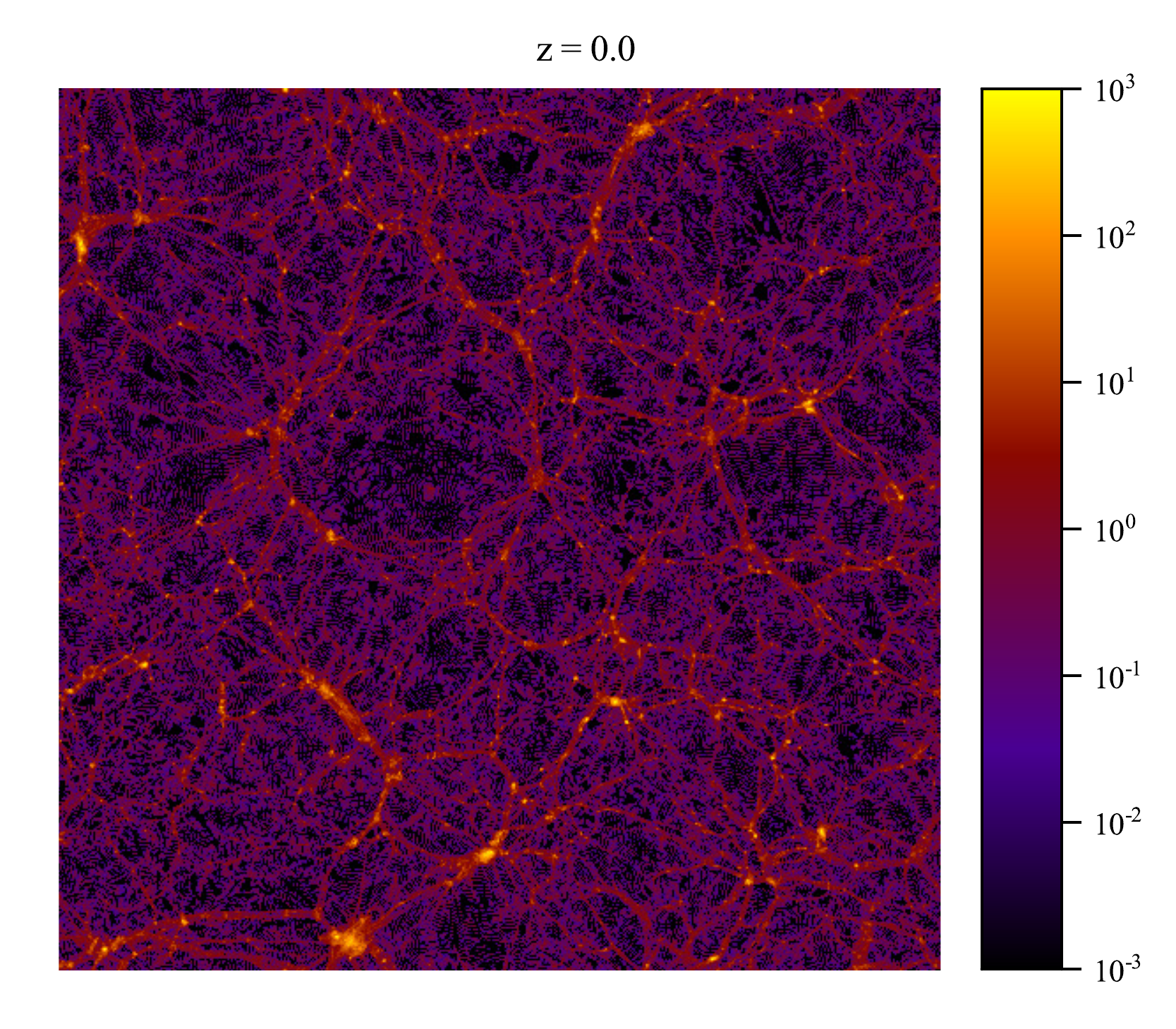}
\includegraphics[width=0.45\textwidth]{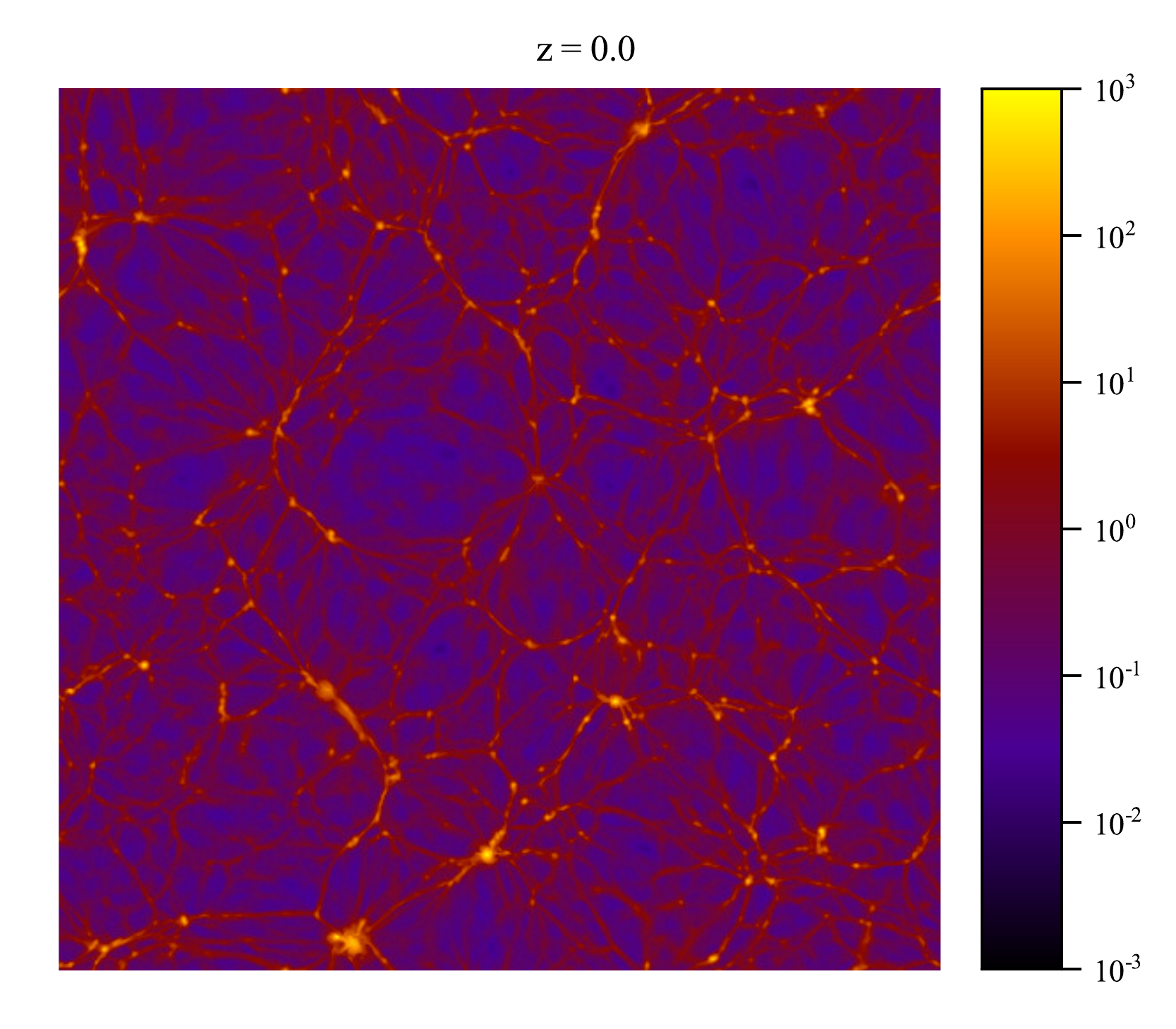}
}
\caption{Two-dimensional slice views of the density distribution of dark matter and baryons. Slices are $100\times100 h^{-2}{\rm Mpc}^2$, with a thickness $0.098\mpch$. The left-hand panels are for dark matter, and the right-hand panels are for baryons. The redshifts are $z=5, 1$ and $0$, arranged from the top to bottom.}
\label{fig:dens2}
\end{figure*}

% --------------------------------------------------------------------------------------
\begin{figure*}
\centerline{
\includegraphics[width=0.45\textwidth]{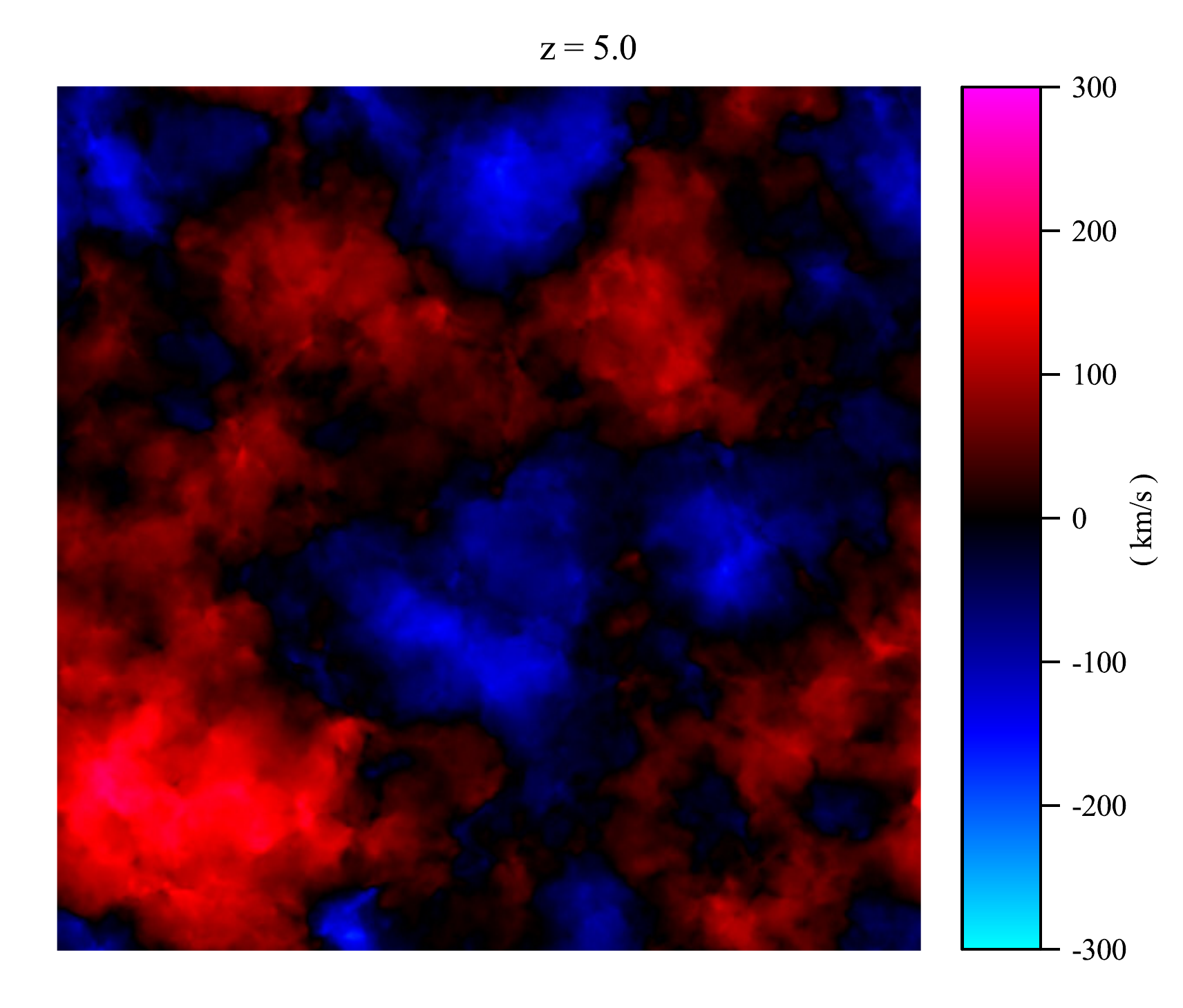}
\includegraphics[width=0.45\textwidth]{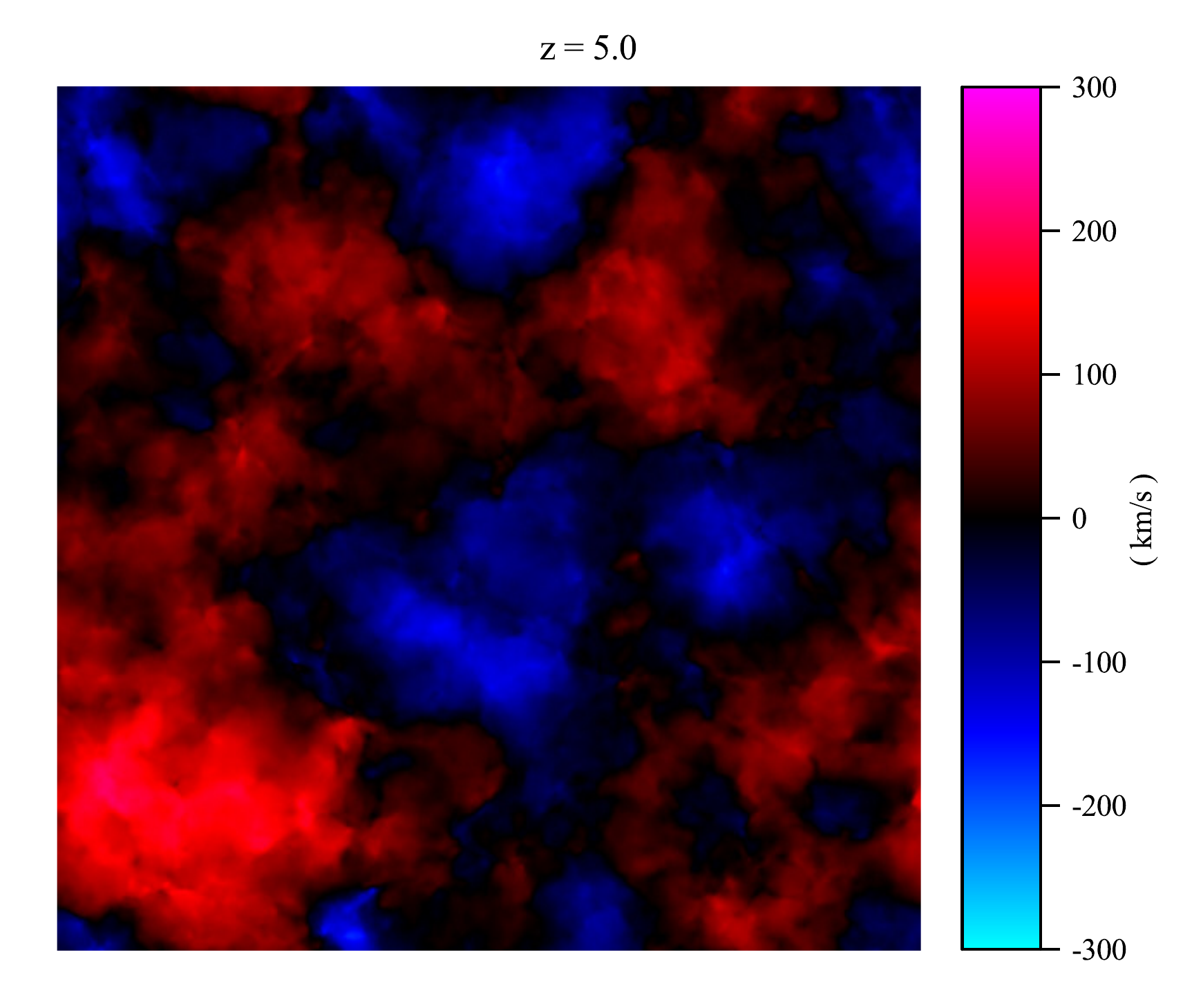}
}
\centerline{
\includegraphics[width=0.45\textwidth]{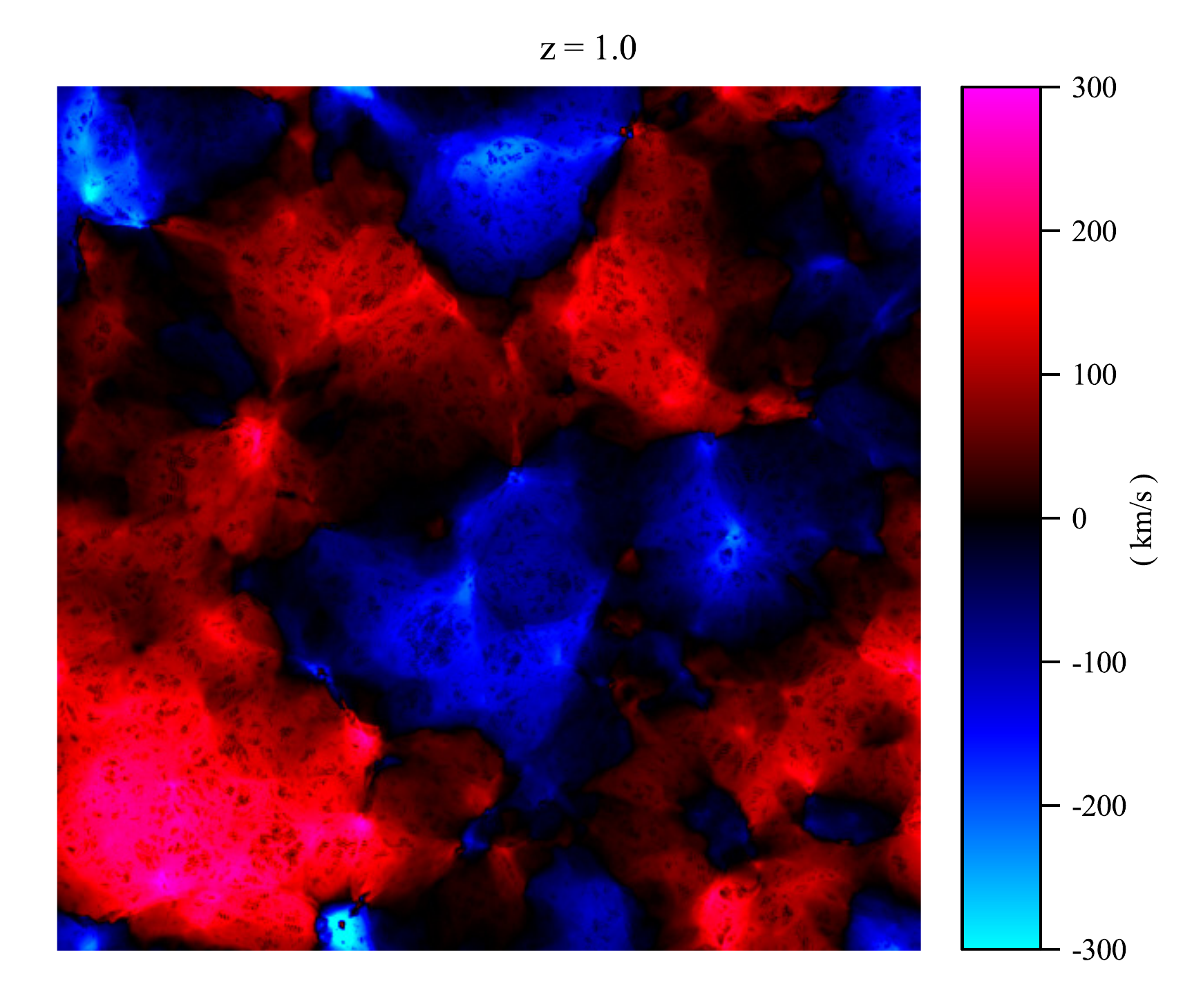}
\includegraphics[width=0.45\textwidth]{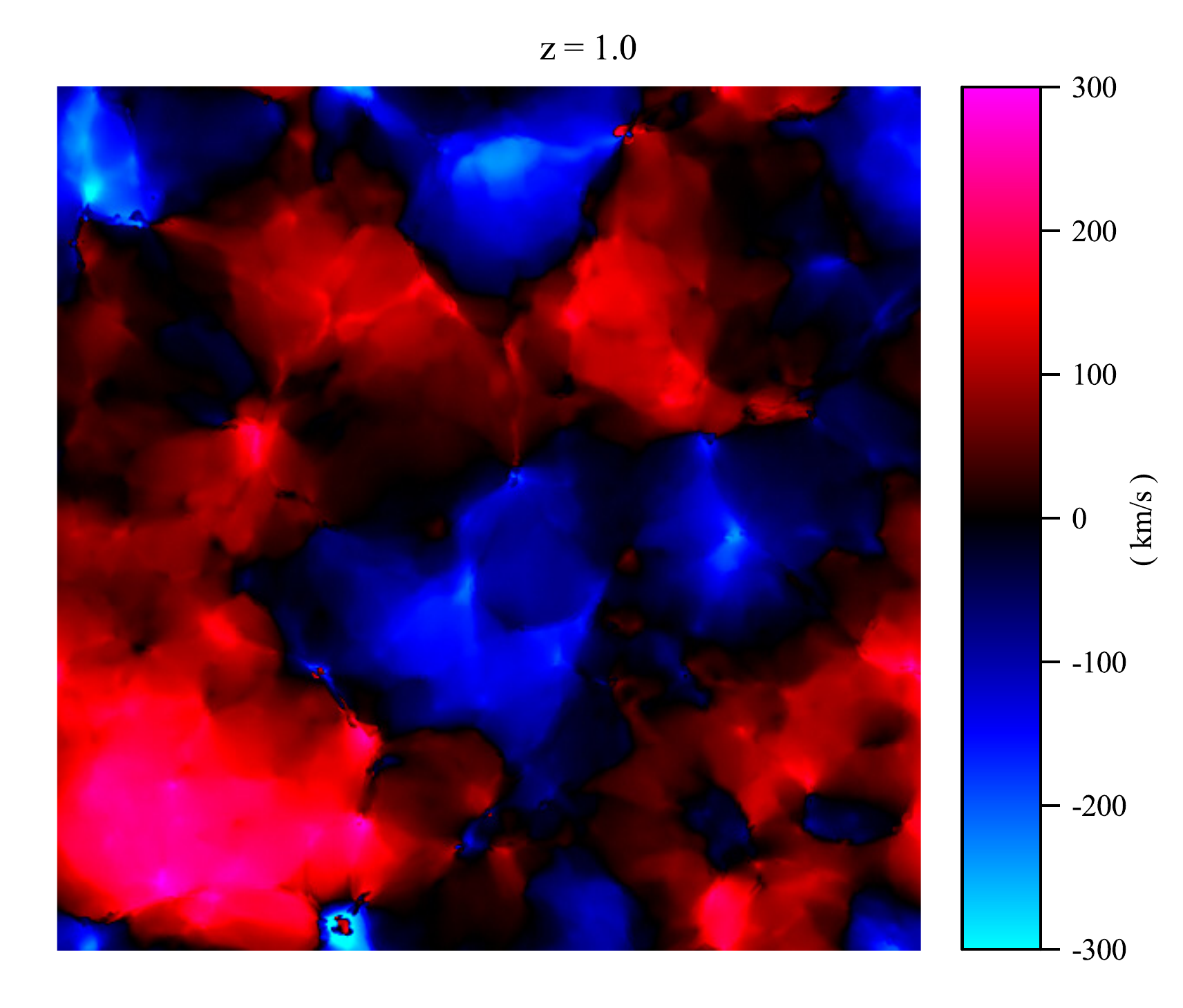}
}
\centerline{
\includegraphics[width=0.45\textwidth]{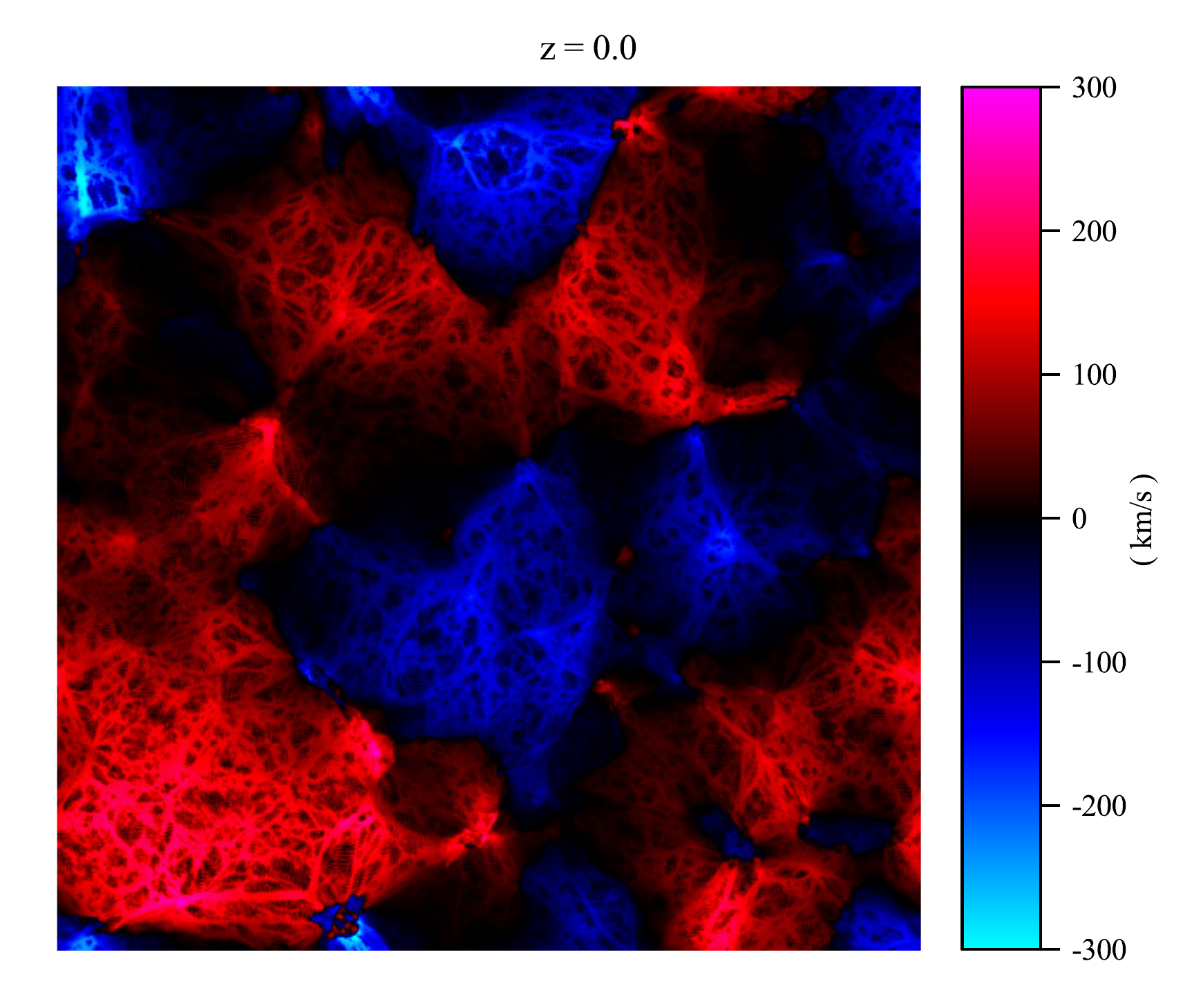}
\includegraphics[width=0.45\textwidth]{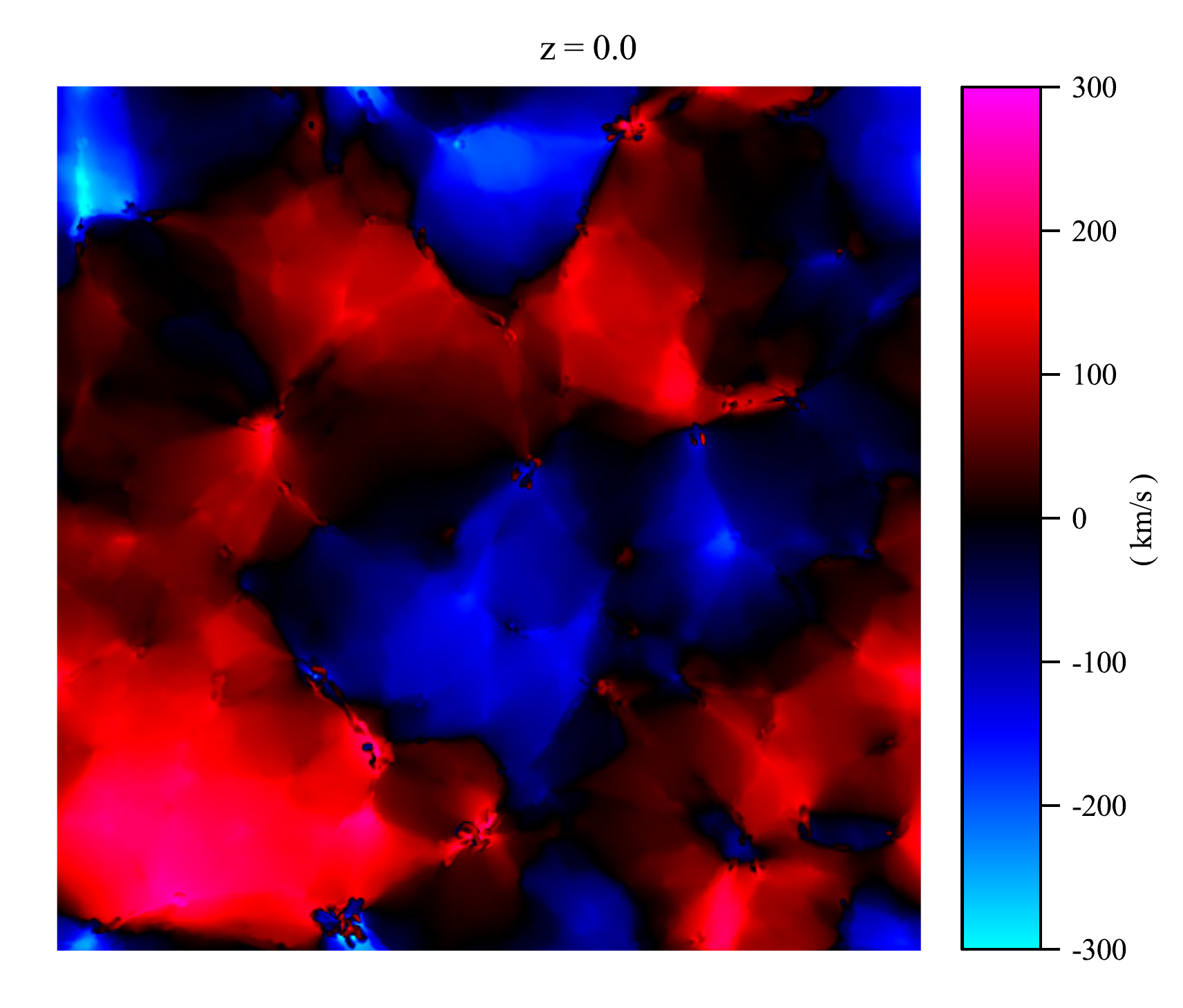}
}
\caption{Two-dimensional slice views of velocity ($z$-direction) distribution of dark matter and baryons. Slices are $100\times100 h^{-2}{\rm Mpc}^2$, with a thickness $0.098\mpch$. The left-hand panels are for dark matter, and the right-hand panels are for baryons. The redshifts are $z=5, 1$ and $0$, arranged from the top to bottom.}
\label{fig:vel2}
\end{figure*}

% --------------------------------------------------------------------------------------
\section{Linear Bias Model}
\label{sec:linear}

If the matter content of the Universe is dominated by cold dark matter (CDM), then for CDM and baryons, the linear evolution of their density contrast, $\delta_\dm(\vk)$ and $\delta_\ba(\vk)$, in Fourier space are as follows \citep{Mo2010}:
% --------------------------------------------------------------------------------------
\begin{eqnarray}
\label{eq:cdm}
\frac{\dd^2\delta_{\rm dm}}{\dd t^2} + 2\frac{\dot{a}}{a} \frac{\dd\delta_{\rm dm}}{\dd t} & = & 4\pi G\bar{\rho}_{\rm m}\delta_{\rm dm}, \nonumber\\
\frac{\dd^2\delta_{\rm b}}{\dd t^2} + 2\frac{\dot{a}}{a} \frac{\dd\delta_{\rm b}}{\dd t} + \frac{v_s^2 k^2}{a^2} \delta_{\rm b}& = & 4\pi G\bar{\rho}_{\rm m}\delta_{\rm dm},
\end{eqnarray}
in which $\bar{\rho}_{\rm m}$ is the mean total matter density and $v_s$ is the sound speed, whose value depends on the temperature of IGM, and hence $v_s$ is generally a function of spatial coordinates. If $v_s$ is assumed to be a constant, and since
% --------------------------------------------------------------------------------------
\begin{displaymath}
\vv(\vk) = \frac{ia{\vk}}{k^2} \frac{\dd\delta({\vk})}{\dd t},
\end{displaymath}
we have the solution of equation~(\ref{eq:cdm}) as
% --------------------------------------------------------------------------------------
\begin{equation}
\label{eq:ldvk}
\delta_\ba(\vk) =  \frac{\delta_\dm(\vk)}{1 + x^2_{\rm J}k^2}, \hspace{20pt}
\vv_\ba(\vk) =  \frac{\vv_\dm(\vk)}{1 + x^2_{\rm J}k^2},
\end{equation}
in which $\delta_\ba(\vk)$, $\delta_\dm(\vk)$, $\vv_\ba(\vk)$, and $\vv_\dm(\vk)$ are the Fourier modes of density contrast and peculiar velocity for baryons (`b') and dark matter (`dm') respectively, and $x_{\rm J} \equiv v_s/2\pi(\pi/G\rho)^{1/2}$ is the Jeans length. For detailed derivations of equation~(\ref{eq:ldvk}), we refer the reader to \citet{Fang1993} and \citet{Bi1993}.

From equation~(\ref{eq:ldvk}), we can see two features: (1) Deviations between baryons and dark matter for density and velocity are in the same fashion, i.e., the two deviations take the same bias factor $1/(1 + x^2_{\rm J}k^2)$; (2) deviations are large at small scales while asymptotically vanishing as the scale gradually goes to infinity, i.e., $k \rightarrow 0$. For example, assume the possible minimum structures as typical dwarf galaxies with mass of several $10^8\msun$, whose Jeans length is roughly estimated with linear theory as $x_{\rm J} = (3M/4\pi\rho_{\rm m})^{1/3}\sim0.1\mpch$, then, according to equation~(\ref{eq:ldvk}), the $1\%$ deviation occurs at the scale $\sim6\mpch$, beyond which the deviation between baryons and dark matter gradually vanishes.

As an application of the linear bias model, \citet{Ma2018} constructed a linear velocity bias model to calculate the cross-correlation function between the kinetic Sunyaev-Zel'dovich effect \citep[kSZ;][]{Sunyaev1972, Sunyaev1980} and the reconstructed peculiar velocity field, in an attempt to constrain the optical depth of galaxies from {\it Planck} data \citep{pc37}. For the modelling, they assume a linear bias between peculiar velocity $v_{\rm g}(k)$ and density contrast $\delta_{\rm m}(k)$ of galaxies as $v_{\rm g}(k) = b_{\rm v}\delta_{\rm m}(k)$, with
% --------------------------------------------------------------------------------------
\begin{equation}
\label{eq:mbias}
b_{\rm v}(k) = 1 + b\left(\frac{k}{k_0}\right)^n,
\end{equation}
in which $k_0$ is a pivot scale, fixed to be $k_0=0.1\hmpc$. With the parameter $n=1.81$ and $b=-0.94$, this model resembles the linear bias model of equation~(\ref{eq:ldvk}) and works well to reproduce the {\it Planck} data. However, when $k>k_0$, the model predicts $b_{\rm v}<0$, which may indicate a loophole for this model.

However, the linear bias model is not a realistic model, since it is impossible for IGM to take a uniform temperature (and hence a constant sound velocity) everywhere, and thus this model is oversimplified. We have to consider a more realistic model to describe the spatial deviation between baryons and dark matter.

% --------------------------------------------------------------------------------------
\begin{figure*}
\centerline{\includegraphics[width=0.85\textwidth]{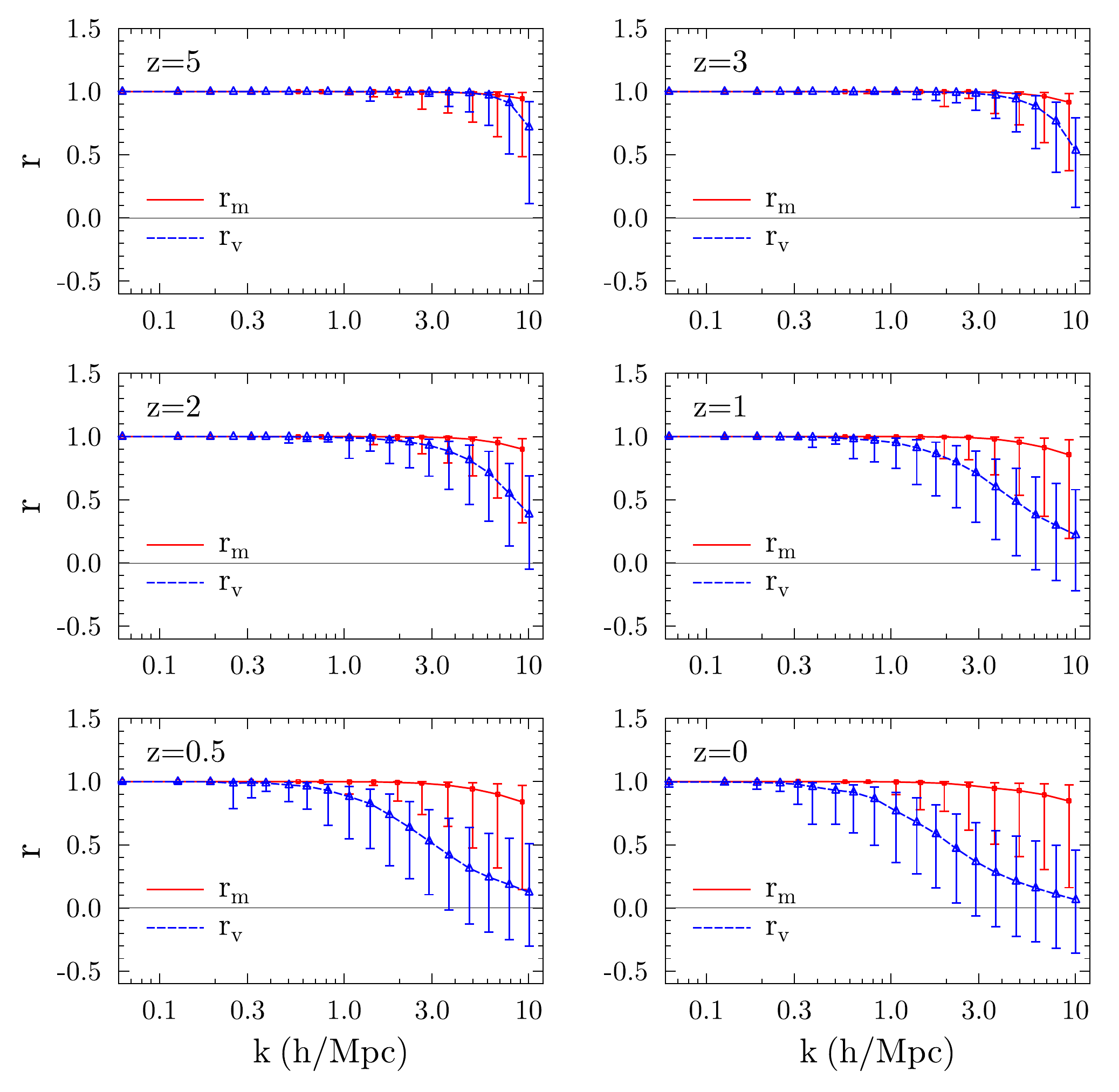}}
\caption{Correlation functions $r_{\rm m}(k)$ and $r_{\rm v}(k)$, as defined in equation~(\ref{eq:rmvk}). Red solid lines are for $r_{\rm m}$, while blue dashed lines are for $r_{\rm v}$. Redshifts from $z=5$ to $0$ are shown in the six panels. The valid scale range is set to be $k<8\hmpc$. See Section~\ref{sec:valscale} for the discussion about the validity scale range. Since the probability distribution of $r(k)$ is highly non-Gaussian and skewed, we use the data below and above the mean value to compute the standard deviation separately, and, in this way, show the lower and upper error in the figure.}
\label{fig:rmvz}
\end{figure*}

% --------------------------------------------------------------------------------------
\section{Simulation and Data}
\label{sec:data}

For the current simulations, we generate the initial density fluctuation from Gaussian random field in Fourier space at redshift $z=99$, of which the linear power spectrum is calculated with the approach of \citet{Eisenstein1999}. Dark matter particles are placed on a uniform cubic gird at first, and then their positions and initial peculiar velocity are set by Zel'dovich approximation. The initial density and velocity of baryons at grid points follow exactly those of dark matter particles, except for the difference of a factor, i.e. the ratio between the two densities.

We refer the interested readers to \citet{Valkenburg2017} for error analyses and discussions about the latest approaches for generating accurate initial conditions in mixed dark matter-baryon simulations.

The simulations in this paper were performed using the hybrid cosmological $N$-body/hydrodynamical code WIGEON, which is based on the positivity-preserving WENO finite-difference scheme for the hydro-solver and incorporated with the standard PM method for the gravity calculation of dark matter particles \citep{Feng2004, Zhu2013}. For the current simulation code, we use the plain PM scheme to compute the gravity between dark matter particles, without including adaptive mesh refinement treatment. Due to the five-order accuracy of its hydro-solver, this code is much effective to capture shockwave and turbulence structures of IGM. The simulations are evolved from the initial time to the present day in periodic cubical boxes with side length 100$\mpch$. With an equal number of grid cells and dark matter particles of $1024^3$, the space and mass resolutions are determined to be 97.7$h^{-1}{\rm kpc}$ and $8.3\times10^7\msun$, respectively. A uniform UV background is switched on at $z = 11$ to mimic the re-ionization. The radiative cooling and heating processes are modelled with a primordial composition (${\rm X} = 0.76, {\rm Y} = 0.24$) following the method in \citet{Theuns1998}. As addressed in Section~\ref{sec:intro}, star formation, SN and AGN feedback effects are not included in the current work.

The set of data were used to investigate the properties of motions of the IGM and the impact of numerical viscosity on turbulence \citep{Zhu2013}, the growth of vortical motions of the baryonic gas and the evolution of mass and velocity field in the cosmic web \citep{Zhu2015,Zhu2017}. In this paper, we use these data to explore the deviation scale between baryons and dark matter through density field and velocity field.

% --------------------------------------------------------------------------------------
\section{Results}
\label{sec:result}

We show two-dimensional slice views of the spatial distribution of density and velocity for dark matter and baryons in Figs.~\ref{fig:dens2} and \ref{fig:vel2}, respectively. We see that, at high redshifts, say $z=5$, the distributions of density and velocity for baryons and dark matter are nearly the same. While as time increases, the deviations of spatial distributions between the two become more and more significant, and at the present become the largest. In the following, we explore the deviations in more details.

% --------------------------------------------------------------------------------------
\subsection{Scale-dependent deviation}
\label{sec:sclbias}

We study the scale-dependent deviation between baryons and dark matter, by constructing the following two dimensionless cross-correlation functions as a function of Fourier modulus $k$, $r_{\rm m}(k)$ and $r_{\rm v}(k)$, for density field (`m') and velocity field (`v'), respectively, as
% --------------------------------------------------------------------------------------
\begin{eqnarray}
\label{eq:rmvk}
r_{\rm m}(k) & = & \langle\frac{\delta_{\rm dm}(\vk)\delta^{*}_{\rm b}(\vk)}{|\delta_{\rm dm}(\vk)| |\delta_{\rm b}(\vk)|}\rangle, \nonumber \\
r_{\rm v}(k) & = & \langle\frac{\vv_{\rm dm}(\vk)\cdot\vv^*_{\rm b}(\vk)}{|\vv_{\rm dm}(\vk)| |\vv_{\rm b}(\vk)|}\rangle,
\end{eqnarray}
in which `$*$' denotes complex conjugate, and `$<...>$' denotes the statistical averaging among all the Fourier modes with modulus $k$. Note that if the spatial deviations between baryons and dark matter are described by the linear bias model of equation~(\ref{eq:ldvk}), then no matter what the form of the bias factor is, $r_{\rm m}(k)$ and $r_{\rm v}(k)$ are always equal to 1, so the two correlation functions defined in equation~(\ref{eq:rmvk}) reflect entirely non-linear spatial deviations between baryons and dark matters.

We apply the fast Fourier transform technique to the simulation data aforementioned in Section~\ref{sec:data} to compute the Fourier modes of density field ($\delta_{\rm dm}(\vk)$ and $\delta_{\rm b}(\vk)$) and velocity field ($\vv_{\rm dm}(\vk)$ and $\vv_{\rm b}(\vk)$), and then according to equation~(\ref{eq:rmvk}) to compute the two correlation functions. From Fig.~\ref{fig:rmvz}, we see that both the correlation functions approach one at $k\rightarrow0$, and decrease when $k$ increases, which indicate that at larger and larger scales, the deviations between baryons and dark matter are vanishing, while at smaller and smaller scales, the deviations gradually become more and more significant.

However, from Fig.~\ref{fig:rmvz}, it is interesting to note that, without exceptions, all the $r_{\rm v}$ curves lie below $r_{\rm m}$, and at $k\sim10\hmpc$, $r_{\rm v}$ can even drop to nearly $0$ at $z=0$. The vanishing $r_{\rm v}$ indicates complete uncorrelation or deviation between $\vv_{\rm b}(\vk)$ and $\vv_{\rm dm}(\vk)$, while $r_{\rm m}\approx0.8-0.9$, indicating density fields are still correlated, or partially deviated, with each other between baryons and dark matter. This result shows that the deviations revealed by velocity field are more significant than that by density field, as is different from the linear bias model of equation~(\ref{eq:ldvk}).

% --------------------------------------------------------------------------------------
\begin{table}
\vspace{5pt}
\begin{center}
\begin{tabular*}{0.425\textwidth}{cccccccc}
\hline\hline
\multirow{2}{*}{$z$} & \multicolumn{3}{c}{$k_{\rm r}(\hmpc)$} & & \multicolumn{3}{c}{$k_{\rm r}(\hmpc)$}\\
\cline{2-4} \cline{6-8}
 $ $ & $r_{\rm m}$=0.99 & 0.95 & 0.90  &  & $r_{\rm v}$=0.99 & 0.95  & 0.90  \\
\hline
  0 &~~~~~~~1.69 & 3.83  & 5.98 &  &~~~~~~0.27  & 0.37 & 0.65 \\
0.5 &~~~~~~~2.40 & 4.65  & 6.75 &  &~~~~~~0.29  & 0.66 & 0.96 \\
  1 &~~~~~~~2.97 & 5.35  & 7.40 &  &~~~~~~0.46  & 1.10 & 1.50 \\
  2 &~~~~~~~3.87 & 6.75  & 9.20 &  &~~~~~~1.13  & 2.39 & 3.49 \\
  3 &~~~~~~~4.39 & 7.54  & 9.75 &  &~~~~~~2.45  & 4.46 & 5.75 \\
  5 &~~~~~~~4.55 & 8.86  & 12.0 &  &~~~~~~4.80  & 6.91 & 8.03 \\
\hline
\end{tabular*}
\caption{The deviation scale $k_{\rm r}$ as a function of $z$ and $r_{\rm m}$ or $r_{\rm v}$, corresponding to Fig.~\ref{fig:krz}.}
\end{center}
\label{tab:tab1}
\end{table}

% --------------------------------------------------------------------------------------
\begin{figure}
\vspace{5pt}
\centerline{\includegraphics[width=0.45\textwidth]{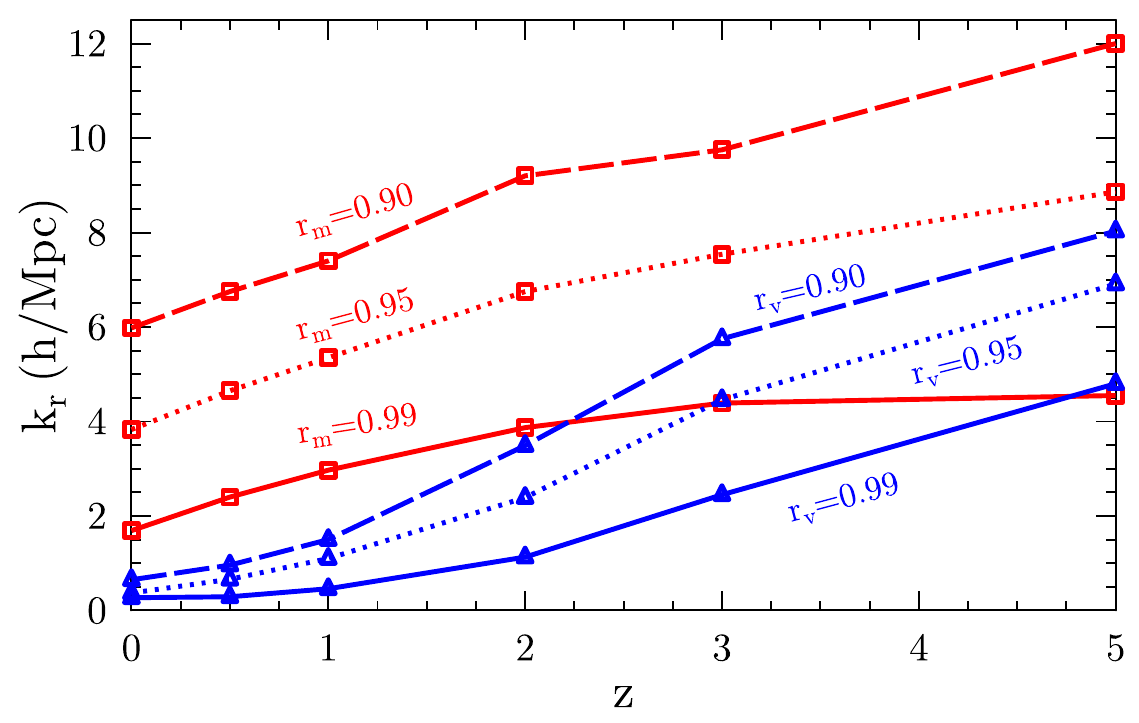}}
\caption{The deviation scale $k_r$ as a function of $z$. The relevant data are listed in Table~\ref{tab:tab1}. Red lines are for $r_{\rm m}$, and blue lines are for $r_{\rm v}$, respectively. The corresponding value of $r_{\rm m}$ and $r_{\rm v}$ are indicated in the figure.}
\label{fig:krz}
\end{figure}

% --------------------------------------------------------------------------------------
\begin{figure*}
\centerline{\includegraphics[width=0.85\textwidth]{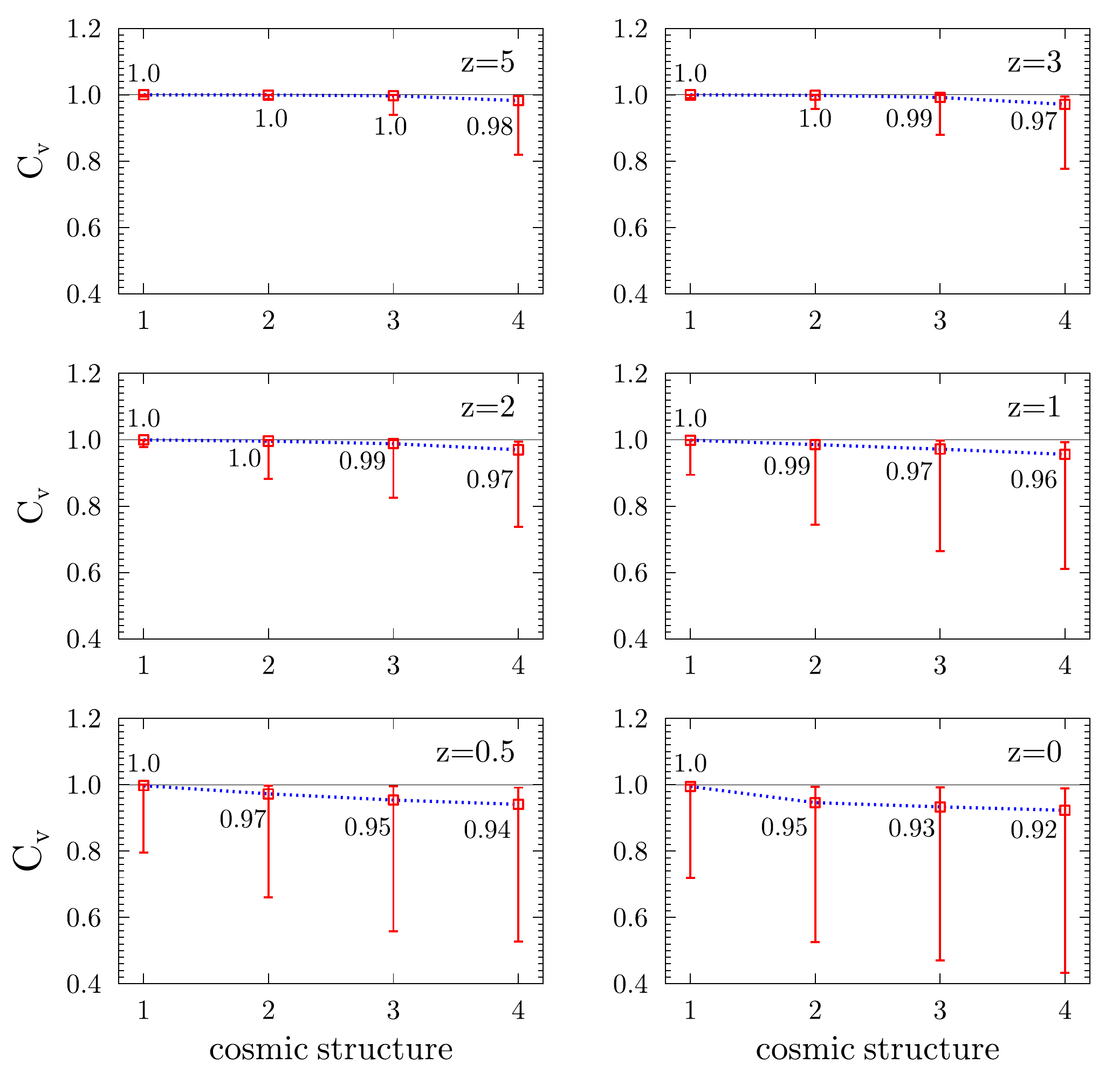}}
\caption{The velocity correlation function defined in equation~(\ref{eq:cv}) as a function of the four cosmic structures, in the order of increasing density from 1 to 4. The four cosmic structures are specified in Section~\ref{sec:envbias} as (1) voids and underdense regions, (2) sheets and filaments, (3) outskirts of clusters, and (4) virialized clusters. The corresponding $C_{\rm v}$ values are indicated in the figure. Redshifts from $z=5$ to $0$ are shown in the six panels. Errors are calculated in the same way as those of Fig.~\ref{fig:rmvz}.}
\label{fig:cvz}
\end{figure*}

% --------------------------------------------------------------------------------------
From the results of Fig.~\ref{fig:rmvz}, corresponding to every $r_{\rm m}$ or $r_{\rm v}$, we can define a deviation scale $k_r$ (or $\lambda_r\equiv2\pi/k_r$). In Table~\ref{tab:tab1}, we list $k_r$ at $r_{\rm m}$ or $r_{\rm v} = 0.99, 0.95$, and $0.90$ at the redshift $z=0,0.5,1,2,3$, and $5$, respectively. In Fig.~\ref{fig:krz}, we show these $k_r$ as a function of $z$, with corresponding values of $r_{\rm m}$ or $r_{\rm v}$ indicated in the figure. From the figure, we can see that the separation or difference between $r_{\rm m}$ or $r_{\rm v}$ is apparent. At $z=0$, for the $1\%$ deviation\footnote{Since $|r_{\rm m}$ or $|r_{\rm v}$ is always $<1$, we tentatively define the {\it degree of deviation} = $1-|r_{\rm m}|$ or $1-|r_{\rm v}|$. Hence $r_{\rm m}$ or $r_{\rm v}=0$ suggests a 100\% deviation.} ($r_{\rm m}=0.99$) of the density field, from Table~\ref{tab:tab1}, we can derive that the deviation scale $\lambda_r =2\pi/k_r = 3.7\mpch$, roughly a scale of a cluster of galaxies. While for the velocity field, the deviation scale for the $1\%$ deviation ($r_{\rm v}=0.99$) can reach as large as $\lambda_r = 23\mpch$, a scale that falls within the weakly non-linear regime for the structure formation paradigm. Also, it is evident that, in all cases, the deviation scale $\lambda_r$ becomes smaller and smaller with increasing redshift. At $z=5$, the $1\%$ deviation scales are $1.4$ and $1.3\mpch$ for density and velocity field, respectively. As mentioned in Section~\ref{sec:intro}, the deviation of per cent level caused by SN and AGN feedback between baryons and dark matter occurs at $k\sim1\hmpc$ ($\lambda\sim6\mpch$). Hence, our results indicate that the effect of turbulence heating is {\it indeed} comparable to that of these feedback processes.

As addressed previously, with the current simulation data, we found that the bias model, equation~(\ref{eq:mbias}) in \citet{Ma2018} is problematic, in which the bias factor is real function, and in this work, we are forced to model the bias as complex function.

The bias model is connected with $r_{\rm m}$ or $r_{\rm v}$ coefficients as follows. Let
% -----------------------------------------------------------------------------------------------
\begin{equation}
\label{eq:bias}
\delta_{\rm b}(\vk) = b(\vk)\delta_{\rm dm}(\vk),
\end{equation}
in which $b(\vk)$ is the bias between $\delta_{\rm b}(\vk)$ and $\delta_{\rm dm}(\vk)$. If $b(\vk)$ is a complex function of $\vk$, as $b(\vk) = |b(\vk)|(\cos\theta_b(\vk) + i \sin\theta_b(\vk))$, then
% --------------------------------------------------------------------------------------
\begin{eqnarray}
\label{eq:rcmplx}
r_{\rm m}(k) & = & \langle\frac{\delta_{\rm dm}(\vk)\delta^*_{\rm b}(\vk)}{|\delta_{\rm dm}(\vk)| |\delta_{\rm b}(\vk)|}\rangle = \langle\frac{b^*(\vk)}{|b(\vk)|}\rangle \nonumber \\
& = & \langle\cos\theta_b(\vk)\rangle - i \langle\sin\theta_b(\vk)\rangle.
\end{eqnarray}
Similar results can also be obtained for $r_{\rm v}(k)$. Hence, $\langle \sin \theta_b \rangle$ is zero statistically, and Fig.~\ref{fig:rmvz} is actually plotted for $\langle\cos\theta_b\rangle$. If $b(\vk)$ is a real function, then no matter what its functional form is, $r_{\rm m}(k)$ or $r_{\rm v}(k)$ are always equal to one.

We further explore the physical effects caused by phase differences between these Fourier modes. According to the convolution theorem, equation~(\ref{eq:bias}) can be translated into the following convolution in real space, as
% ----------------------------------------------------------------------------------------------
\begin{equation}
\label{eq:densconv}
\delta_{\rm b}({\bf x}) = \int b({\bf x} - {\bf x}')\delta_{\rm dm}({\bf x}')\dd{\bf x}',
\end{equation}
where $\delta_{\rm b}({\bf x})$ and $\delta_{\rm dm}({\bf x})$ are density fields for baryons and dark matter, respectively, and $b({\bf x})$ is a spatial function, whose Fourier transform is $b({\vk})$. Since when $\vk$ goes to zero, the bias is vanishing, i.e. $b(\vk)\rightarrow 1$, so if $b(\vk)$ is a real function, then its real-space counterpart $b({\bf x})$ is a symmetric function, and can be treated as a smoothing function to make a smoother density field of baryons from dark matter fields. In this case, the locations of density peaks for baryons are the same as those for dark matter peaks. While if $b(\vk)$ is a complex function, then $b({\bf x})$ is asymmetric, which cannot be considered as a smoothing function, and hence the locations of density peaks of baryons may be different from those of dark matter.

Additionally, Fig.~\ref{fig:rmvz} also shows that the bias $b(\vk)$ is a function of $z$, which suggests that the motions of baryons and dark matter are not {\it synchronous} due to the time dependence of the bias; thus, the spatial distributions of baryons and dark matter are very complicated.

In Section~\ref{sec:vdev}, we discuss the possible dynamics of the spatial separations of both density and velocity fields between baryons and dark matter.

% --------------------------------------------------------------------------------------
\begin{figure}
\vspace{5pt}
\centerline{\includegraphics[width=0.45\textwidth]{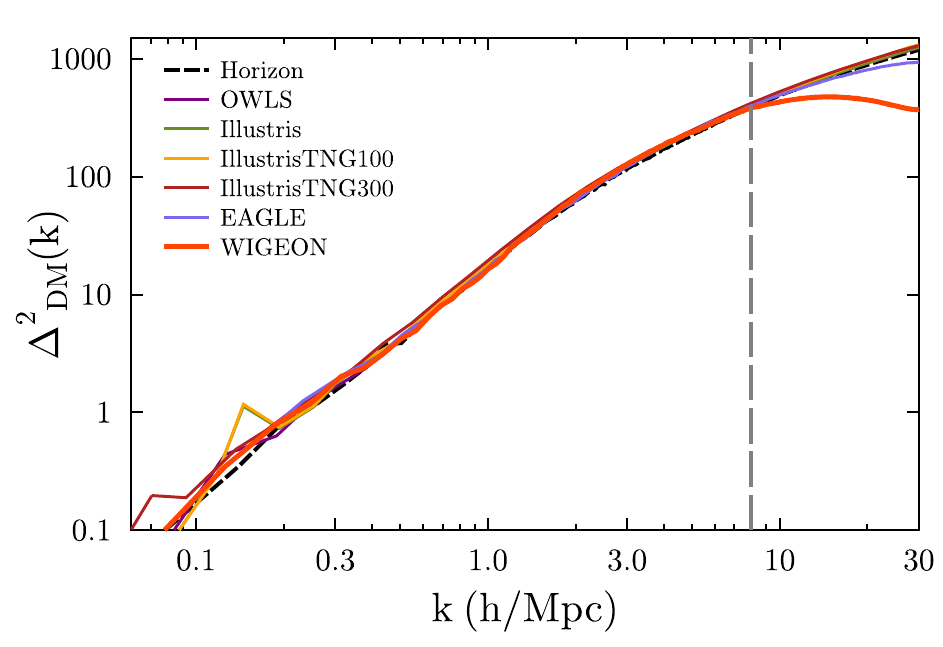}}
\caption{Comparison of the dark-matter-only power spectrum at $z=0$ between our simulation (WIGEON) and Horizon \citep{Chisari2018}, OWLS \citep{vanDaalen2011}, Illustris \citep{Vogelsberger2014}, IllustrisTNG \citep{Springel2018}, and EAGLE \citep{Hellwing2016}. In the figure, $\Delta^2(k) = k^3P(k)/2\pi^2$. The left part of the vertical dashed line at $k=8\hmpc$ is the validity scale range that we discussed in Section~\ref{sec:valscale}.}
\label{fig:deltak_ps}
\end{figure}

% --------------------------------------------------------------------------------------
\subsection{Environment-dependent deviation}
\label{sec:envbias}

The two correlation functions, $r_{\rm m}(k)$ and $r_{\rm v}(k)$ of equation~(\ref{eq:rmvk}), derived from Fourier transform, reflect scale-dependent deviations of the spatial distribution between baryons and dark matter. It is well known that Fourier transform can just provide information in Fourier space, and all information in real space is smeared out. However, one may be interested in how the results revealed by $r_{\rm m}(k)$ and $r_{\rm v}(k)$ correspond to the relevant cosmic structures in real space. For this purpose, following the scheme of \citet{Zhu2013}, we first consider some cosmic structures classified by the total matter density as follows:
% --------------------------------------------------------------------------------------
\begin{enumerate}
\item $0<\rho_t/\rho_{\rm crit}<6$  :  voids and underdense regions,
\item $6<\rho_t/\rho_{\rm crit}<36$ :  sheets and filaments,
\item $36<\rho_t/\rho_{\rm crit}<200$ :  outskirts of clusters,
\item $200<\rho_t/\rho_{\rm crit}<\infty$ : virialized clusters,
\end{enumerate}
where $\rho_t$ ($=\rho_{\rm b} + \rho_{\rm dm}$) , $\rho_{\rm b}$, $\rho_{\rm dm}$, and $\rho_{\rm crit}$ are the total, baryon, dark matter, and critical density, respectively. The density boundaries, $6$, $36$, and $200$, are slightly different from those in \citet{Vazza2009} and are identical to the mean density of sheets, filaments, and halos predicted in \citet{Shen2006}.

Based on the above classification of cosmic structures, we define a velocity cross-correlation function $C_{\rm v}(\rho)$ between baryons and dark matter, which is a function of matter density as:
% --------------------------------------------------------------------------------------
\begin{equation}
\label{eq:cv}
C_{\rm v}(\rho)=\langle\frac{\vv_{\rm b}(\rho)\cdot\vv_{\rm dm}(\rho)}{|\vv_{\rm b}(\rho)||\vv_{\rm dm}(\rho)|}\rangle,
\end{equation}
in which $\vv_{\rm b}(\rho)$ and $\vv_{\rm dm}(\rho)$ via matter density $\rho$ implicitly depend on spatial coordinates, and `$<...>$' indicates averaging over all spatial points with the same $\rho$. For a cosmic structure of some kind with the density $\rho$, $C_{\rm v}(\rho)$ through velocity field reflects the deviation between baryons and dark matter. In practice, however, we treat $C_{\rm v}$ as a function of density interval as $C_{\rm v}(\rho_1<\rho_t<\rho_2)$, in which $\rho_1$ and $\rho_2$ are the density boundaries to specify the four cosmic structures abovementioned.

In Fig.~\ref{fig:cvz}, we show the results of $C_{\rm v}$ as a function of four cosmic structures. We can see that the deviations between baryons and dark matter are the most prominent for the densest structures (virialized clusters), then diminish for the looser and looser structures, in the order of outskirts of clusters, sheets, and filaments, and voids and underdense regions. Also, these deviations evolve with redshifts -- they become more and more prominent with decreasing redshifts. Although $C_{\rm v}$ cannot provide accurate spatial positioning for the deviations, in combining the results of $r_{\rm m}(k)$ or $r_{\rm v}(k)$, we see that the most significant deviations occur both at large $k$'s (small scales) and meanwhile at densest structures. Hence, the results revealed by $C_{\rm v}$ and by $r_{\rm m}(k)$ or $r_{\rm v}(k)$ are consistent with and complementary to each other.

% --------------------------------------------------------------------------------------
\begin{figure*}
\centerline{\includegraphics[width=0.85\textwidth]{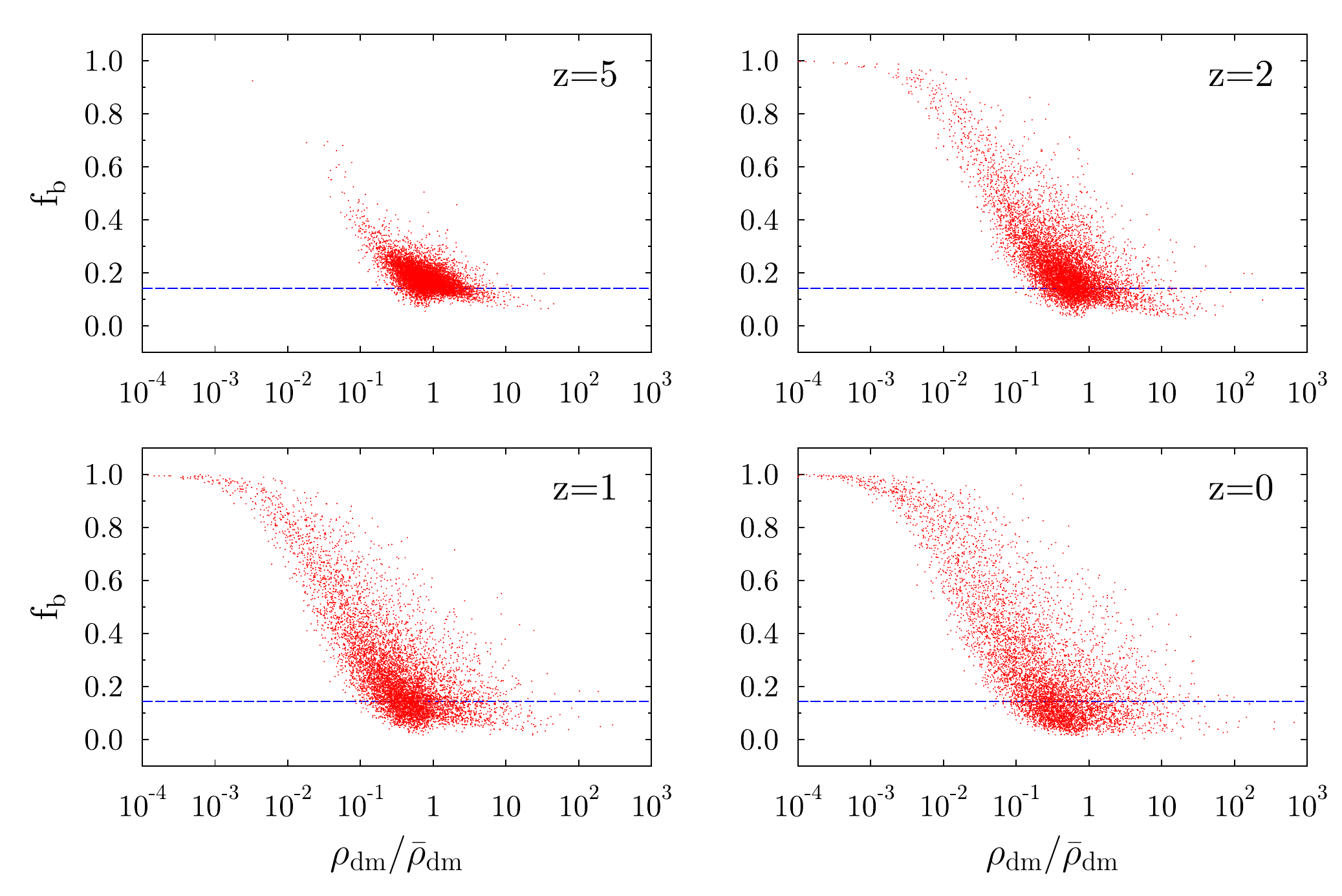}}
\caption{Scatter plot of baryon fraction $f_{\rm b}$ vs. dark matter density $\rho_{\rm dm}$, divided by the cosmic mean value $\bar{\rho}_{\rm dm}$, from the simulation samples at redshifts $5, 2, 1$, and $0$. The blue dashed line indicates the cosmic mean value of baryon fraction, $\bar{f}_{\rm b} = 0.143$.}
\label{fig:bfrac}
\end{figure*}

% --------------------------------------------------------------------------------------
\subsection{Validity scale range}
\label{sec:valscale}

One may ask a question that within what scale range our results can be trusted. We here present an analysis for the valid scale range.

As addressed in Section~\ref{sec:data}, the simulations were performed using the cosmological $N$-body/hydrodynamical scheme, which is based on the positivity-preserving WENO finite-difference scheme for the hydro-solver and incorporated with the PM method for the gravity calculation of dark matter particles. The hydro-solver is very accurate because of its five-order accuracy algorithm and hence, compared with the gravity calculation of dark matter particles, errors introduced by hydro-solver can be neglected. Therefore, we only concentrate on the gravity calculation of dark matter particles. In order to determine the validity scale range, we perform a dark-matter-only simulation and compare our dark matter power spectrum with others' high-resolution results \citep{vanDaalen2011, Vogelsberger2014, Hellwing2016, Chisari2018, Springel2018}. From Fig.~\ref{fig:deltak_ps}, we see that our result agrees well with others' at the left part of the vertical line at $k=8\hmpc$, indicating that our PM algorithm is reliable down to the scale of $8\hmpc$ at $z=0$. Hence, based on the above analyses, we choose $k < 8\hmpc$ as the validity scale at $z = 0$. Our results in Section~\ref{sec:sclbias} should be trusted within this scale range. At higher redshifts, the validity scale range should be larger than that at $z=0$.

% --------------------------------------------------------------------------------------
\begin{figure*}
\centerline{\includegraphics[width=0.85\textwidth]{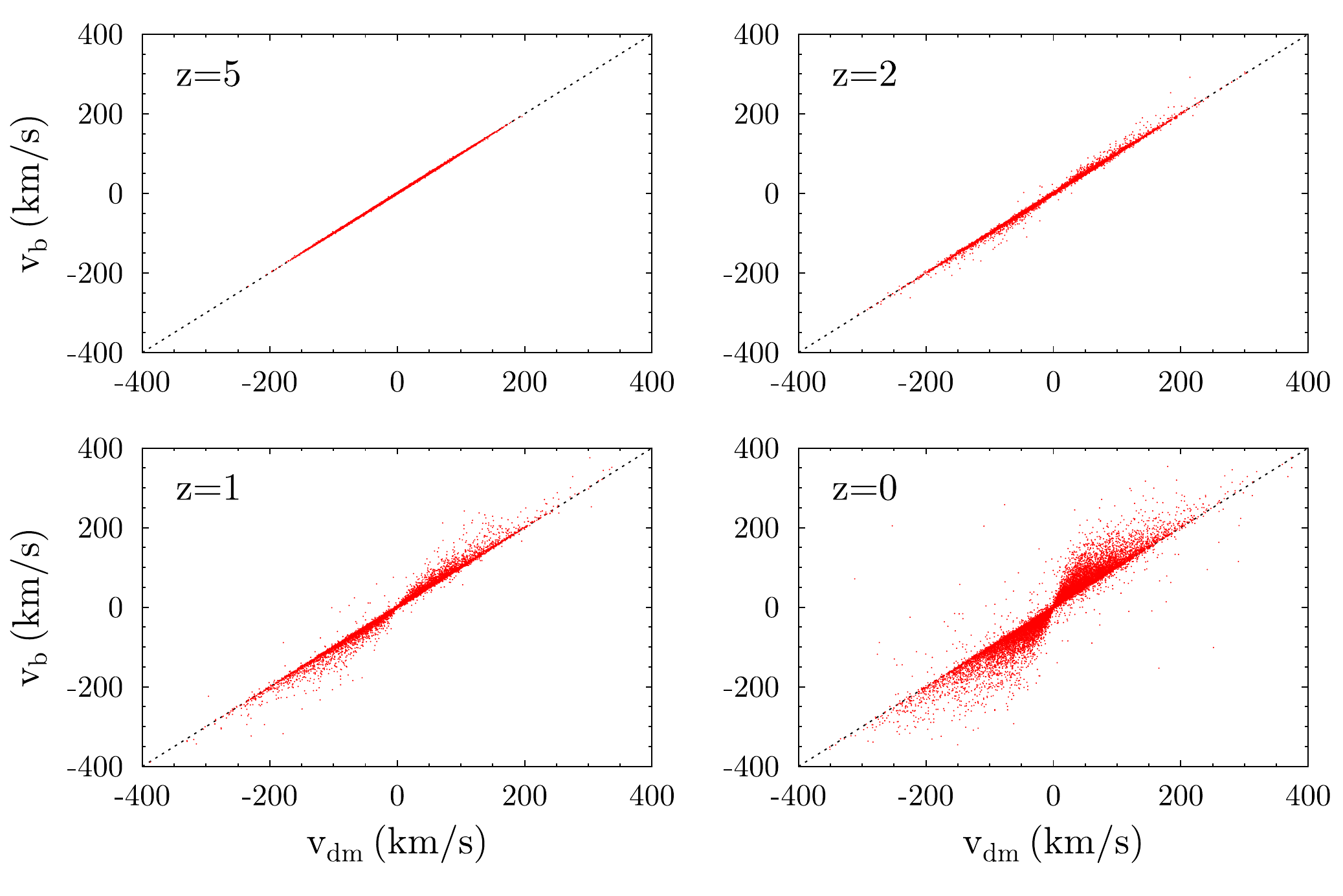}}
\caption{Scatter plot of $v_{\rm b}$ vs. $v_{\rm dm}$ from the simulation samples at redshifts $5, 2, 1$, and $0$. For simplicity, only $z$-direction velocity is plotted. The black dotted line indicates $v_{\rm b} = v_{\rm dm}$.}
\label{fig:vv2}
\end{figure*}

% --------------------------------------------------------------------------------------
\subsection{Mechanism for density and velocity deviation}
\label{sec:vdev}

In a series of previous works, we have established a preliminary framework of turbulence heating for cosmic baryonic fluid. We revealed that the highly evolved cosmic baryonic fluid at low redshifts can be characterized by the She–Leveque scaling law, which implies that the cosmic baryonic fluid is in a state similar to a fully developed turbulence within the scale range $\lambda < 3\hmpc$ \citep{Hep2006, Zhu2010}. We also investigated the properties of motions of the IGM and the impact of numerical viscosity on turbulence \citep{Zhu2013}, the growth of vortical motions of the baryonic gas, and the evolution of mass and velocity field in the cosmic web \citep{Zhu2015, Zhu2017}.

Different from the heating mechanisms by SN or AGN feedback, the turbulence heating for IGM may be simply addressed as follows. During gravitational collapsing of IGM, turbulence will emerge in IGM when Reynolds number is sufficiently high, and energy from gravitational potential energy of baryonic gas is converted into kinetic energy, and then cascades to smaller and smaller scales by turbulence and eventually dissipates, re-heating the gas.

Based on turbulence heating for IGM, we attempt to present the mechanism for the spatial deviations between baryons and dark matter of density and velocity field, and also explain why the deviation scale derived from the velocity field is larger than from the density field.

Any vector field, say a velocity field $\vv$, can be separated by the Helmholtz-Hodge decomposition \citep{Arfken2005} into divergence (or longitudinal) part and curl (or transverse) part, as $\vv = \vv_{\rm div} + \vv_{\rm curl}$, in which the divergence part $\vv_{\rm div}$ satisfies $\nabla\times\vv_{\rm div}=0$, and the curl part $\vv_{\rm curl}$ satisfies $\nabla\cdot\vv_{\rm curl}=0$, respectively.

For a cosmic baryonic fluid, the dynamical equation of vorticity $\vec{\omega} \equiv \nabla \times \vv = \nabla \times \vv_{\rm curl}$ can be derived from the Euler equation \citep{Zhu2010}:
% --------------------------------------------------------------------------------------
\begin{eqnarray}
\label{eq:dvor}
\frac{D \vec {\omega}}{Dt}& \equiv  & \partial_t {\vec \omega} + \frac{1}{a}{\vv} \cdot {\nabla} \vec{\omega} \nonumber \\
& = & \frac{1}{a}({\bf S} \cdot {\vec\omega} - d {\vec\omega} + \frac{1}{\rho^2} \nabla \rho \times \nabla p-\dot{a}\vec{\omega}),
\end{eqnarray}
in which $p$ is the pressure of the IGM, $a(t)$ is the cosmic scale factor, $d=\nabla\cdot\vv$, $S_{ij} = (1/2)(\partial_iv_j + \partial_jv_i)$, and hence $[{\bf S}\cdot {\vec\omega}]_i = S_{ij} \omega_j$. Obviously, in the linear regime, only the last term of equation~(\ref{eq:dvor}) survives. This term is from the cosmic expansion and makes the vorticity decaying as $a^{-1}$. Thus, the vorticity of the IGM is reasonably negligible in the linear regime.

Equation~(\ref{eq:dvor}) shows that if the initial vorticity is zero, the vorticity will stay at zero in the non-linear regime, provided that the baroclinity term $(1/\rho^2)\nabla \rho \times\nabla p$ is zero. If baryonic gas stays in the thermal equilibrium state with the equation of state $p=p(\rho)$, then $\nabla p$ would be parallel to $\nabla \rho$, and then $(1/\rho^2) \nabla \rho \times\nabla p=0$. However, once multistreaming and turbulent flows have developed \citep{Hep2006, Fang2011}, complex structures, like curved shocks, will lead to a deviation of the direction of $\nabla p$ from that of $\nabla \rho$. In this case, the $\rho-p$ relation cannot be simply given by an single-variable function as $\rho=\rho(p)$, and the baroclinity will no longer be zero. \citet{Zhu2010} show that vorticity field of IGM significantly increases with time, and hence the curl velocity $\vv_{\rm curl}$ also increases with time. In fact, not just in baryonic fluid, \citet{Zhu2017} indicate that vorticity is also significant for the dark matter velocity field, albeit the mechanisms of generating vorticity are different for the two matter.

At early times, the IGM is comoving with dark matter, and all its velocities are almost divergence velocities. As time goes by, however, several effects affect the divergence velocity of IGM. One is from the pressure detainment. In high-density regions, at first IGM falls with dark matter towards the gravitational centres, but later, $\vv_{\rm div}$ of IGM is slowed down by $\nabla p$\footnote{One can consider $\nabla p$ as a buoyancy effect, which always resists the gravitational collapsing.}, while dark matter is not affected by $\nabla p$. So baryons are detained by $\nabla p$ and thus more extended in space. This is one way that baryons and dark matter are separated in spatial distribution.

Another effect comes from the shockwave deceleration. According to shockwave theory, the ratios of the pre- and post-shock density and velocity are as follows \citep{Landau1959, Shu1992}:
% --------------------------------------------------------------------------------------
\begin{equation}
\label{eq:shvv}
\frac{\rho_2}{\rho_1}=\frac{v_1}{v_2}=\frac{(\gamma + 1)Ma^2_1}{2 + (\gamma - 1)Ma^2_1},
\end{equation}
where the subscript `1' and `2' indicates the pre- and post-shock quantities of the baryonic gas, respectively, $Ma$ is the Mach number, and $\gamma$ is the adiabatic index, with $\gamma=5/3$ for single-atom gas. Usually, Mach number is very large for IGM; hence, from equation~(\ref{eq:shvv}), we can see that shocks always lead to deceleration of the $\vv_{\rm div}$ and detainment of baryons during the process of collapsing. However, dark matter is not affected by shocks, and thus baryons and dark matter are separated in spatial distribution.

In Fig.~\ref{fig:bfrac}, we plot the baryon fraction $f_{\rm b} (=\rho_{\rm b}/\rho_{\rm t})$ as a function of dark matter density $\rho_{\rm dm}$ and redshift $z$. At $z=5$, we see that most of the baryons are located within dark matter regions of $0.1<\rho_{\rm dm}/\bar{\rho}_{\rm dm}<10$, and the baryon fractions are concentrated at about the cosmic mean value of $\bar{f}_{\rm b}=0.143$. As analysed above, as time increases, baryons are more and more extended in space. As a result, baryon fractions are relatively low at high-density regions but are high in low-density regions, as can be seen in the panels from $z=2$ to $0$.

% --------------------------------------------------------------------------------------
\begin{figure}
\vspace{5pt}
\centerline{\includegraphics[width=0.45\textwidth]{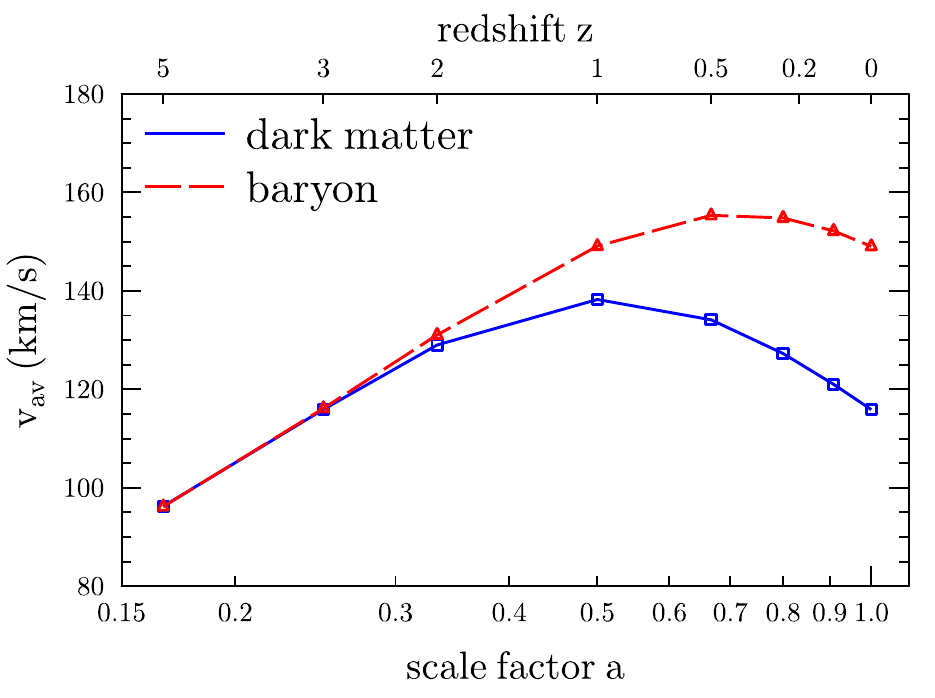}}
\caption{The averaged velocity of dark matter (blue line) and baryonic matter (red) as a function of the scale factor $a$. The averaged velocity $v_{\rm av}$ for baryons and dark matter are calculated over all the grid points.}
\label{fig:averv2}
\end{figure}

From Fig.~\ref{fig:vv2}, we see that at first, baryons are comoving with dark matter, so that there is nearly no velocity difference between them (the panel of $z=5$). While as times goes by, the two velocities are more and more deviated from each other, with baryon velocity more and more increased, as shown by panels from $z=2$ to $0$. This can also be seen from Fig.~\ref{fig:averv2}. At the early time ($z=5$), the averaged velocity of baryons is the same as that of dark matter. Due to dynamical evolution, both averaged velocities increase at the same pace. At about $z=1$, for the dark matter, the cosmological expansion\footnote{Without dynamics, the cosmological expansion always decelerates the matter's peculiar velocity $v$ with the scale factor $a$ as $v \propto a^{-1}$.} exceeds the velocity enhancement by dynamical effect, so that the velocity of dark matter begins to decrease when $z<1$. For baryons, the time evolution of the averaged velocity roughly has the same trend as that of dark matter, but increases faster than that of dark matter, due to the more and more increasing of the curl velocity $\vv_{\rm curl}$.

In the above, we just provide some simplified explanations for the spatial deviations between baryons and dark matter. Moreover, there are also some other effects, such as star wind, SN feedback, or AGN feedback, affect the velocity of IGM, but since these effects are not included in our simulations, we do not discuss them here.

Furthermore, the divergence part and curl part of IGM's velocity affect the spatial distribution of baryons in different ways. From the continuous equation,
% --------------------------------------------------------------------------------------
\begin{eqnarray}
\label{eq:cons}
\frac{\partial\rho}{\partial t} + \frac{1}{a} {\nabla} {\cdot}(\rho\vv) & \equiv & \frac{D\rho}{D t} + \frac{\rho}{a}\nabla\cdot\vv \nonumber \\
& = & \frac{D\rho}{D t} + \frac{\rho}{a}\nabla\cdot \vv_{\rm div} = 0,
\end{eqnarray}
in which $D\rho/Dt\equiv\partial\rho/\partial t + a^{-1}\vv\cdot\nabla\rho$. We can see that the streamlines of baryonic gas are determined by $\vv$, but the time evolution of baryon density along the streamline is affected just by the divergence velocity $\vv_{\rm div}$, not by curl velocity $\vv_{\rm curl}$. Due to the difference in the time evolution of baryon density, $\vv_{\rm div}$ and $\vv_{\rm curl}$ undoubtedly affect the spatial distribution of baryons in different ways.

To summarize, together with all the above analyses, we have the following: (1) The pressure gradient $\nabla p$ and shockwave effects from the divergence part of baryon velocity tend to prevent baryons from falling into gravitational centres, so that the spatial distribution of baryons is more extended than that of dark matter at later times; (2) the curl velocities of baryons tend to increase at a faster pace than the divergence velocities with increasing time, but do not affect the time evolution of baryons; and, as a result, (3) the deviation scale of baryons from dark matter derived by velocity field is larger than that by density field, as shown in Figs.~\ref{fig:rmvz} and \ref{fig:krz}.

% ------------------------------------------------------------------------------------------------
\section{Summary and Conclusions}
\label{sec:concl}

Both theories and observations of modern cosmology show that baryons and dark matter do not distribute in the same way in space. That is, the spatial distributions between the two matter components of the Universe are deviated or biased from each other. The question is what kind of mechanisms causes these deviations? Simulations, such as OWLS, Illustris, IllustrisTNG, Horizon, and EAGLE, indicate that feedback processes like galactic winds from star formation and SNe, especially AGN activity, play pivotal roles to separate baryons from dark matter in spatial distribution. In this work, however, we would like to explore how turbulence heating will influence motions and spatial distributions of IGM. For this purpose, we investigate on what scale the deviation of spatial distributions between baryons and dark matter occurs, by using numerical simulation samples produced by the hybrid cosmological hydrodynamical/$N$-body code. For the code, we use the plain PM scheme to solve the gravity of dark matter particles, and use the WENO scheme, which is effective for capturing shockwave and turbulence structures in the baryonic gas, to solve the hydrodynamical equations.

Our series of previous works have already revealed that the highly evolved cosmic baryonic fluid at low redshifts can be characterized by the She–Leveque scaling law, which implies that the cosmic baryonic fluid is in a state similar to a fully developed turbulence within the scale range $\lambda < 3 \mpch$. In this work, we do not include the subgrid processes such as star formation, SN and AGN feedback into our simulation code. In this way the effect of turbulence heating can be revealed to the most extent. We summarize our findings as follows:

(1) In equation~(\ref{eq:rmvk}), we construct two cross-correlation functions $r_{\rm m}(k)$ and $r_{\rm v}(k)$ for density field and velocity field, respectively. As indicted by Fig.~\ref{fig:rmvz}, the deviations will become larger and larger if $r_{\rm m}(k)$ and $r_{\rm v}(k)$ are smaller and smaller. The two correlation functions entirely reflect non-linear deviations between baryons and dark matters, since if the deviations are described by the linear bias model of equation~(\ref{eq:ldvk}), then no matter what the functional form of the bias is, $r_{\rm m}(k)$ and $r_{\rm v}(k)$ are always equal to 1.

(2) Our investigations show that both the correlation functions $r_{\rm m}(k)$ and $r_{\rm v}(k)$ approach 1 at $k\rightarrow0$, while decrease with increasing $k$. This result indicates that at larger and larger scales, the deviations between baryons and dark matter are asymptotically vanishing, but at smaller and smaller scales, the deviations gradually become more and more significant. Our results also show that linear bias models of equations~(\ref{eq:ldvk}) or (\ref{eq:mbias}) are poor models to describe the deviations between baryons and dark matter. We demonstrate that the bias function $b(\vk)$ should not be real but be complex function. Otherwise, the deviation between baryons and dark matter cannot be properly described. Fig.~\ref{fig:rmvz} also shows that the bias is a function of redshift, which suggests that the motions of baryons and dark matter are not synchronous due to the time dependence of the bias. Thus, the spatial distributions of baryons and dark matter are very complicated.

(3) Our computations show that, in all cases, the $r_{\rm v}$ curves lie below the $r_{\rm m}$, which indicates that the spatial deviations between baryons and dark matter revealed by velocity field are more significant than that by density field. At $z=0$, for the $1\%$ deviation ($r_{\rm m}=0.99$) of the density field, the deviation scale $\lambda_r = 3.7\mpch$, while for the velocity field, the deviation scale for the $1\%$ deviation ($r_{\rm v}=0.99$) can reach as large as $\lambda_r = 23\mpch$, a scale that falls within the weakly non-linear regime for the structure formation paradigm. Also, it is obvious to see that, in all cases, the deviation scale $\lambda_r$ becomes smaller and smaller with increasing redshift. At $z=5$, the $1\%$ deviation scales are $1.4$ and $1.3\mpch$ for density and velocity field, respectively.

(4) The velocity cross-correlation function $C_{\rm v}$ between baryons and dark matter reveal that the deviations between baryons and dark matter are the most prominent for the densest structures, such as virialized clusters, then diminish for the looser and looser structures, such as outskirts of clusters, sheets and filaments, and voids and underdense regions. Also, these deviations evolve with redshifts; that is, they become more and more prominent with decreasing redshifts. Although $C_{\rm v}$ cannot provide accurate spatial positioning for the deviations, in combining the results of $r_{\rm m}$ or $r_{\rm v}$, we see that the most significant deviations occur both at large $k$'s (small scales) and meanwhile at densest structures. The results revealed by $C_{\rm v}$ are consistent with and complementary to those by $r_{\rm m}$ or $r_{\rm v}$.

(5) The deviation scale between baryons and dark matter derived from the velocity field is larger than that from the density field. As analysed in Section~\ref{sec:vdev}, reasons for the difference between the two deviations should be as follows: (1) The divergence part of baryon velocity tends to prevent baryons from falling into gravitational centres, so that the spatial distribution of baryons is more extended than that of dark matter at later times; (2) the curl velocity does not affect the time evolution of baryons; (3) the curl part of baryon velocity tends to increase at a faster pace than the divergence part with increasing time; and, as a result, (4) the deviation scale of baryons from dark matter derived by velocity field is larger than that by density field.

These results indicate that the effect of turbulence heating for IGM is comparable to that of these processes such as star formation, SN and AGN feedback. Moreover, our result is consistent with \citet{Zhuravleva2014}, who found that turbulence heating is sufficient to offset radiative cooling and appears to balance it locally at each core radius of the two clusters. It seems that turbulence heating should be an important ingredient for the physics of IGM, which is ignored in the current mainstream cosmological simulations.

What dynamical mechanisms can excite turbulence in fluids? Generally speaking, when the Reynolds number is sufficiently large, turbulence will probably occur in the fluid. Some dynamical processes may provide the velocity and possibly change the density of the fluid, and thereby ensure that the Reynolds number reaches sufficiently high values. In the cosmological context, there may be mainly two processes that will excite turbulence in IGM. The first is the structure formation or mergers in an expanding background, as is addressed in Section~\ref{sec:vdev}. The second process is outflows of baryonic gas driven by AGN feedback, which may be an even greater source of eventual turbulence than halo formation/mergers, particularly in galaxy groups and clusters. These issues should be interesting topics for further investigations.

Finally, we address that the large-scale peculiar velocity field of the Universe is a powerful probe of cosmology, in that it directly responds to the gravitational action of all matter and energy and is sensitive to density inhomogeneities, making it a promising tool to study the dark Universe \citep{Ma2015, Zhang2015}. The observations of spatial deviations between baryons and dark matter from both density and velocity field can help support $\Lambda$CDM cosmology and eliminate the possibility of modified gravitational theories (say MOND) as a substitution of dark matter. Recently, various velocity surveys have been conducted, such as Cosmicflows-I/II/III \citep{Tully2008, Tully2013, Tully2016}, and the peculiar velocity-related observations, such as kSZ effect \citep{Sunyaev1972, Sunyaev1980}, have also been performed \citep{Hand2012, Ferraro2016, Hill2016, pc13, Schaan2016, Soergel2016, Bernardis2017}. However, due to the spatial deviations between baryons and dark matter, as revealed by the density and velocity field, cautions should be paid for any dynamical information drawn from velocity observations.

% ------------------------------------------------------------------------------------------------
\section*{Acknowledgments}
We are very grateful to the anonymous referee for many constructive suggestions and comments. The simulations were run at Supercomputing Center of the Chinese Academy of Sciences, and SYSU. PH acknowledges the support by the National Science Foundation of China (No. 11273013), and by the Natural Science Foundation of Jilin Province, China (No. 20180101228JC).

\section*{Data availability}

The data used in this paper are available from the correspondence author upon reasonable request.

% ------------------------------------------------------------------------------------------------

\end{document}